\documentclass[iop]{emulateapj}
\usepackage[version=3]{mhchem}
\usepackage{longtable}
\usepackage{apjfonts,natbib}

\makeatletter
\newcounter{reaction}
\renewcommand\thereaction{C\,\arabic{reaction}}
\newcommand\reactiontag{\refstepcounter{reaction}\tag{\thereaction}}
\newcommand\reaction@[2][]{\begin{equation}\ce{#2}%
\ifx\@empty#1\@empty\else\label{#1}\fi%
\reactiontag\end{equation}}
\newcommand\reaction@nonumber[1]{\begin{equation*}\ce{#1}%
\end{equation*}}
\newcommand\reaction{\@ifstar{\reaction@nonumber}{\reaction@}}
\makeatother

\shorttitle{Helium Atmosphere on Exoplanet}

\shortauthors{Hu et al.}

\begin{document}

\title{Helium Atmospheres on Warm Neptune- and Sub-Neptune-Sized Exoplanets and Applications to GJ~436~b}

\author{Renyu Hu$^{1,2,3}$, Sara Seager$^{4}$, Yuk L. Yung$^{1,2}$}
\affil{$^1$Jet Propulsion Laboratory, California Institute of Technology, Pasadena, CA 91109, USA}
\affil{$^2$Division of Geological and Planetary Sciences, California Institute of Technology, Pasadena, CA 91125}
\affil{$^3$Hubble Fellow}
\affil{$^4$Department of Earth, Atmospheric and Planetary Sciences, Massachusetts Institute of Technology, Cambridge, MA 02139, USA)}
\email{renyu.hu@jpl.nasa.gov \\ Copyright 2015. All rights reserved.}

\begin{abstract}
Warm Neptune- and sub-Neptune-sized exoplanets in orbits smaller than Mercury's are thought to have experienced extensive atmospheric evolution. Here we propose that a potential outcome of this atmospheric evolution is the formation of helium-dominated atmospheres. The hydrodynamic escape rates of Neptune- and sub-Neptune-sized exoplanets are comparable to the diffusion-limited escape rate of hydrogen, and therefore the escape is heavily affected by diffusive separation between hydrogen and helium. A helium atmosphere can thus be formed -- from a primordial hydrogen-helium atmosphere -- via atmospheric hydrodynamic escape from the planet. The helium atmosphere has very different abundances of major carbon and oxygen species from those of a hydrogen atmosphere, leading to distinctive transmission and thermal emission spectral features. In particular, the hypothesis of a helium-dominated atmosphere can explain the thermal emission spectrum of GJ~436~b, a warm Neptune-sized exoplanet, while also consistent with the transmission spectrum. This model atmosphere contains trace amounts of hydrogen, carbon, and oxygen, with the predominance of \ce{CO} over \ce{CH4} as the main form of carbon. With our atmospheric evolution model, we find that if the mass of the initial atmosphere envelope is $10^{-3}$ planetary mass, hydrodynamic escape can reduce the hydrogen abundance in the atmosphere by several orders of magnitude in $\sim10$ billion years. Observations of exoplanet transits may thus detect signatures of helium atmospheres and probe the evolutionary history of small exoplanets.
\end{abstract}

\keywords{ radiative transfer --- atmospheric effects --- planetary systems --- techniques: spectroscopic --- planets and satellites: individual (GJ 436 b)}

\section{Introduction}

Recent exoplanet surveys have discovered warm Neptune- and sub-Neptune-sized planets in tightly bound orbits around their parent stars  \citep{Butler2004,Fressin2013,Howard2013}. Were these planets in our Solar System, their orbits would be interior to that of Mercury. The atmospheres of Neptune-sized exoplanets are assumed to be similar to those of giant planets in our Solar System, i.e., dominated by hydrogen and helium \citep{Rogers2010,Nettelmann2010,Madhusudhan2011,Line2014}. However, this assumption may not be valid for warm extrasolar Neptunes in close-in orbits. Due to stellar irradiation, these planets may have experienced significant atmospheric evolution \citep{Lopez2012,Owen2013}. 

\begin{figure*}
\begin{center}
\includegraphics[width=1.0\textwidth]{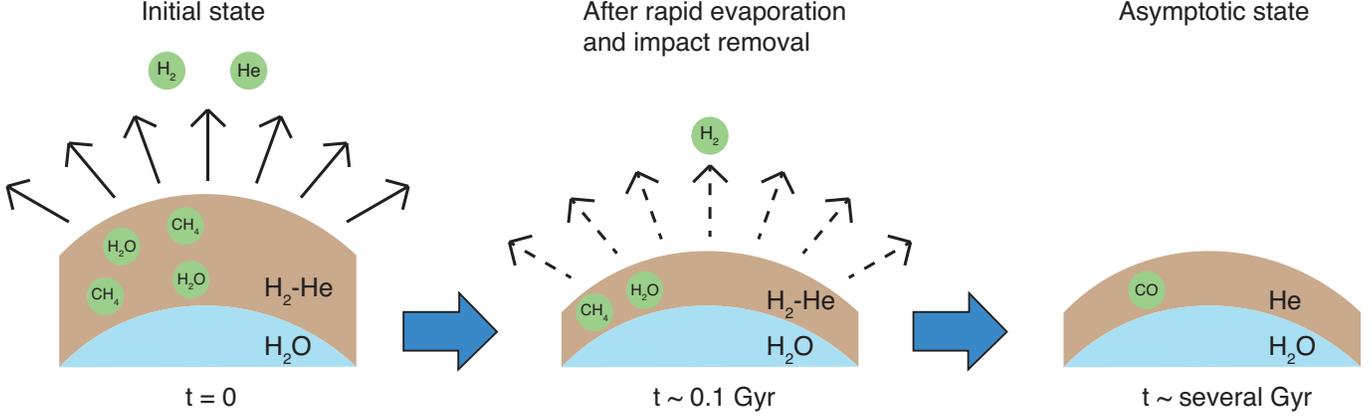}
 \caption{
An evolution scenario for an irradiated Neptune- or sub-Neptune-sized exoplanet that produces a helium-dominated atmosphere from a primordial hydrogen-helium atmosphere. Rapid evaporation could last for $\sim0.1$ Gyr, depending on the thermal history of the planet and the evolution of the stellar X-ray and EUV radiation, and the subsequent thermal escape modulated by diffusive separation between hydrogen and helium could last for multiple billion years to form a helium-dominated atmosphere. The required time depends on the mass of the initial atmosphere.
  }
 \label{Schematic}
  \end{center}
\end{figure*}

Here we consider if some Neptune- and sub-Neptune-sized exoplanets can have their atmospheres depleted in hydrogen but abundant in helium, due to extensive atmosphere loss. Figure \ref{Schematic} illustrates a potential evolution scenario for an exoplanet that leads to an atmosphere dominated by helium. The planet starts with a hydrogen-helium atmosphere accreted from the planet-forming nebula. This initial atmosphere may evaporate and reduce its mass by accretion heating, impact erosion, stellar winds, and X-ray and extreme ultraviolet (EUV) radiation from the parent star. Subsequently, the planet may continue to experience atmospheric loss due to stellar irradiation \citep{Lammer2003,Erkaev2007,Baraffe2008,Lopez2012,Owen2013,Inamdar2014}. If the escape rate is very high, the planet may lose both hydrogen and helium completely. However, if the escape rate is comparable to the diffusion-limited escape rate of hydrogen, the escaping flow may be highly enriched in hydrogen that is 4 times lighter than helium. We show later in this paper that Neptune- and sub-Neptune-sized exoplanets undergoing transonic hydrodynamic escape would generally have an escape rate comparable to the diffusion-limited escape rate of hydrogen, and the fraction of hydrogen in their atmospheres would decrease as the planets evolve. Our model suggests that if the initial mass of the atmosphere is $10^{-3}$ planetary mass, the remaining atmosphere would become helium-dominated within $\sim10$ billion years.

This concept of fossil helium atmosphere can be important because sub-Neptune-sized planets are ubiquitous in our interstellar neighborhood \citep[e.g.][]{Howard2013}, and because helium atmospheres would have very different molecular compositions from hydrogen atmospheres, leading to distinctive spectral features in transmission and thermal emission. The chemical compositions and spectral features of the atmospheres on super Earths and mini Neptunes are highly sensitive to the elemental abundances of the atmospheres, in particular to the hydrogen abundance and the carbon to oxygen (C/O) ratio \citep{Moses2013a,Hu2014}. Compared to other non-\ce{H2}-dominated exoplanet atmospheres, a helium atmosphere has a more extended scale height, and presents a better opportunity to characterize highly evolved exoplanet atmospheres via transmission spectroscopy.

The paper is organized as follows. We describe our models in \S~2. We develop an atmosphere evolution model to study the conditions to form helium atmospheres on Neptune- and sub-Neptune-sized exoplanets, and upgrade an existing atmospheric chemistry and radiative transfer model for exoplanets \citep{Hu2012,Hu2013,Hu2014} to include the capability of treating helium atmospheres. We present a general result of the fractionation between hydrogen and helium for exoplanets undergoing transonic hydrodynamic escape in \S~3. We then apply our hypothesis to the Neptune-sized exoplanet GJ~436~b in \S~4, and propose that the planet can have a helium atmosphere, which explains its puzzling emission features observed with {\it Spitzer}. We discuss how transmission spectroscopy of exoplanets can distinguish highly evolved, helium-dominated atmospheres from hydrogen atmospheres in \S~5, and we conclude in \S~6. 

\section{Model}

\subsection{Atmosphere Evolution Model}

We have created an atmosphere evolution model to investigate whether a primordial hydrogen-helium atmosphere on short-period exoplanets can evolve to a helium-dominated one. We trace the atmosphere mass and elemental abundances from the present time to the past, focusing on the effects of atmosphere escape. This way, our model can be used to test if a helium atmosphere is a plausible evolutionary outcome for an exoplanet. We truncate the model at an age of 0.1 billion years, because the processes dominating the period less than 0.1 billion years after the planet formed (e.g., accretion heating, impact erosion, rapid photoevaporation) do not fractionate hydrogen versus helium. To simplify the calculation, we assume that the core to be unchanged in mass, size, and composition in the evolution history, and that the atmosphere is isothermal. The core could cool down and shrink by up to 1 R$_{\oplus}$ over the history \citep{Baraffe2008}. We have verified that the evolution history of the atmospheric hydrogen abundance is not qualitatively sensitive to these assumptions.

A key element of the evolution model is escape of hydrogen and helium from the envelope. For irradiated exoplanets, the escape is mainly driven by the stellar EUV radiation \citep[e.g.][]{Lammer2003, Yelle2004, Lecavelier2004, Tian2005, GarciaMunoz2007, Murray-Clay2009, Guo2013, Koskinen2013, Lammer2013}, and the escape rate can be estimated by an energy-limited escape formula that relates the EUV irradiation to adiabatic expansion and cooling of the outflow, with an efficiency ($\eta$) that umbrellas the underlying complex energetic and hydrodynamic processes \citep[e.g.][]{Erkaev2007}. Typical values for $\eta$ vary from a few percent to a few tens percent and may change during the evolution depending on the composition and the ionization degree of the flow \citep{Murray-Clay2009,Owen2013}. It has also been suggested that X-ray irradiation may drive escape very early in the history when the stellar X-ray luminosity is high \citep{CecchiPestellini2006,CecchiPestellini2009,Owen2012}. However, X-ray-driven escape is less important for our purpose than EUV-driven escape because much of the X-ray energy deposition is reradiated away and does not drive escape \citep{Owen2012}, and because the X-ray flux is one-order-of magnitude smaller than the EUV flux after 1 billion years since formation (see below).

The energy-limited escape rate is
\begin{equation}
\Phi_{\rm EL} = \frac{L_{\rm EUV} \eta a^2 r_p^3}{4 K d^2 G M_p}, \label{TotalFlux}
\end{equation}
in which $L_{\rm EUV}$ is the luminosity of the star in EUV, $\eta$ is the heating efficiency, or the fraction of absorbed EUV energy that drives escape, $a$ is the ratio between the planet's EUV absorbing radius and the planetary radius, $r_p$ is the planetary radius, $M_p$ is the planetary mass, $d$ is the semi-major axis, $G$ is the gravitational constant, and $K$ is the potential energy reduction factor due to the Roche lobe effect \citep{Erkaev2007}. For convenience we define $Q$ as the incident EUV heating rate on a planet, i.e.
\begin{equation}
Q=\frac{L_{\rm EUV} r_p^2}{4 d^2 },\label{Q}
\end{equation}
and $Q_{\rm net}=\eta Q$ as the net EUV heating rate that drives escape. 

For $L_{\rm EUV}$, we can adopt the following empirical estimates for stellar EUV fluxes based on X-ray observations and coronal modeling of 80 stars with spectral types spanning from M to F \citep{Sanz-Forcada2011}
\begin{eqnarray}
&&\log(L_{\rm EUV}) = 22.12-1.24\log(\tau),\label{EUV}\\
&&\log(L_{\rm X-ray}) = 21.28-1.44\log(\tau),\label{XRAY}
\end{eqnarray}
where the luminosity has a unit of J s$^{-1}$ and the age ($\tau$) has a unit of billion year. This estimate of the EUV luminosity has an uncertainty of an order of magnitude for any individual star; however, we will see later that our result does not sensitively rely on the EUV luminosity. Equations (\ref{EUV}) and (\ref{XRAY}) indicate that the X-ray luminosity can only be a minor contribution to the escape energy for the bulk part of the evolution.

The planetary radius is defined to be the radius of the homopause, i.e., the level above which the binary diffusion coefficient between hydrogen and helium is greater than the eddy diffusion coefficient. Such a definition is only nominal because significant hydrogen photochemistry may occur, and the mixing ratio of \ce{H} is not a constant below the homopause \citep[e.g.][]{Liang2003,Yelle2004,Hu2012,Koskinen2013}. For an eddy diffusion coefficient ranging in $10^9\sim10^{11}$ cm$^2$ s$^{-1}$ \citep{Parmentier2013}, the homopause is at $10^{-3}\sim10^{-5}$ Pascal for a temperature of 600 K, and $10^{-1}\sim10^{-3}$ Pascal for a temperature of $10^4$ K.

Most of the EUV irradiation is deposited at the $\tau=1$ pressure level \citep{Yung1999},
\begin{equation}
P_{1} = n_{1}kT \sim \frac{1}{\sigma H}kT \sim \frac{m_{\ce{H}}g}{\sigma},
\end{equation}
where $H$ is the scale height, $m_{\ce{H}}$ is the mass of hydrogen, $g$ is the gravitational acceleration, and $\sigma=2\sim6\times10^{-18}$ cm$^2$ is the cross section of hydrogen for the EUV irradiation in 13.6-20 eV \citep{Spitzer1978}. For GJ~436~b ($g\sim10$ m s$^{-2}$), the pressure level is $3\sim8\times10^{-5}$ Pascal. Therefore, the EUV absorbing altitude is up to 9 scale heights above the homopause, which corresponds to less than 0.1 planetary radii. In other words, $a<1.1$. We therefore assume $a=1$ in this work,  effectively assuming the uncertainty in $a^2$ in Equation (\ref{TotalFlux}) is absorbed in a wide range of $\eta$. The parameter $a$ being close to unity is also verified by full thermosphere models \citep[e.g.][]{Yelle2004,Murray-Clay2009,Koskinen2013}. We note that this assumption does not imply EUV heating occurs at the homopause, and the parameter $a$ can be greater for smaller planets \citep[e.g.][]{Lammer2013}.

In order to apply Equation (\ref{TotalFlux}) to calculate the escape rate, we need to determine the regime for the hydrodynamic outflow. Hydrodynamic outflow from planetary atmospheres powered by external heating has recently been calculated by Monte Carlo direct simulations, and the advantage of these simulations is that they no longer require outer boundary conditions and that they self-consistently determine the location of the Knudsen layer between the static and the hydrodynamic parts of the atmosphere \citep{Tucker2012, Erwin2013, Johnson2013, Volkov2013}. These simulations find the energy-limited escape rate to be a good approximation of the escape rate when the escape flow is subsonic, but not when the escape flow becomes transonic. The minimum heating rate to drive a transonic outflow is \citep{Johnson2013}
\begin{equation}
Q_{\rm net} > Q_c \sim 4\pi r_* \frac{\gamma}{c_c\sigma_c Kn_m} \sqrt{\frac{2U(r_*)}{m}}U(r_p), \label{Tran_con}
\end{equation}
where $Q_c$ is the critical heating rate, $r_*$ is the sonic point radius, $\gamma$ is the heat capacity ratio, $c_c\sigma_c$ is the collisional cross section, $Kn_m$ is the maximum Knudsen number for the flow to be in the continuum regime, $U$ is the gravitational energy of the escaping particle, and $m$ is the mass of the particle. For an atomic flow, we adopt $\gamma=5/3$, $c_c\sigma_c=5\times10^{-20}$ m$^2$, $Kn_m\sim1$ (valid when the heat is primarily absorbed over a broad range of radius below $r_*$), and $r_*\sim r_0$. Note that $r_*$ can be greater than $r_p$, but $Q_c$ only weakly depends on its specific value as $r_*^{1/2}$. 

In the transonic escape regime, the escape rate no longer increases with the energy put; instead, incremental energy input would be taken away as thermal and translational energy \citep{Johnson2013}. Therefore, Equation (\ref{TotalFlux}) can significantly overestimate the escape rate for transonic flow. We instead estimate the escape rate as
\begin{equation}
\Phi = f_r\Phi_{\rm EL}, \label{TotalFluxR}
\end{equation}
where $f_r$, adopted from \cite{Johnson2013}, is the escape rate reduction factor due to the flow being transonic. $f_r$ is a function of $Q_{\rm net}/Q_c$: when $Q_{\rm net}/Q_c<1$, $f_r\sim1$ and we return to the conventional energy-limited escape; however, when $Q_{\rm net}/Q_c>1$, $f_r<1$ and it decreases nearly as inverse-proportionally to a greater $Q_{\rm net}/Q_c$. 

Equation (\ref{TotalFluxR}) allows us to trace the history of atmosphere loss in terms of the total mass loss flux. Next, in order to trace the effect of atmosphere loss on the atmosphere's elemental abundance, we need to partition the flux into the mass flux of escaping hydrogen and the mass flux of escaping helium, i.e.,
\begin{equation}
\Phi = \Phi_\ce{H} + \Phi_\ce{He}  = 4\pi r_p^2 (\phi_\ce{H}m_\ce{H} + \phi_\ce{He}m_\ce{He}),\label{CompoFlux}
\end{equation}
where $\phi_\ce{H}$ and $\phi_\ce{He}$ denote the number fluxes, and $m_\ce{H}$ and $m_\ce{He}$ denote the masses of hydrogen and helium atom, respectively. The escaping flow should be made of atomic hydrogen and helium, because photochemistry models show that the dominant form of hydrogen in the upper atmosphere should be atomic hydrogen, due to photodissociation of hydrogen molecules catalyzed by water vapor \citep{Liang2003,Yelle2004,Moses2011,Hu2012}.

The physical reason for different $\Phi_\ce{H}$ and $\Phi_\ce{He}$ is that the helium atom is 4 times more massive than hydrogen and that both fluids are subject to the gravity of the planet. When $\Phi$ is large, the coupling between the two components is strong and therefore the two fluxes are approximately proportional to the mixing ratios of the gases at the homopause, which results in little fractionation. When $\Phi$ is small enough helium will not escape and hydrogen escape will be in the so-called ``diffusion-limited'' escape regime. The nature of mass fractionation in hydrodynamic escape has been investigated thoroughly with hydrodynamic formulations \citep{Zahnle1986a, Hunten1987, Zahnle1990}. 

Two complications arise as we apply the classical calculation of mass fractionation to irradiated exoplanets. We have shown earlier that the homopause can be quite close to the EUV absorption radius, which means that (1) the temperature of the homopause can be much higher than the planetary equilibrium temperature due to EUV heating; and (2) the outflow can be partially ionized and the ion-neutral interactions should be considered in assessing whether escaping hydrogen can drag helium.

\begin{figure}
\begin{center}
 \includegraphics[width=0.5\textwidth]{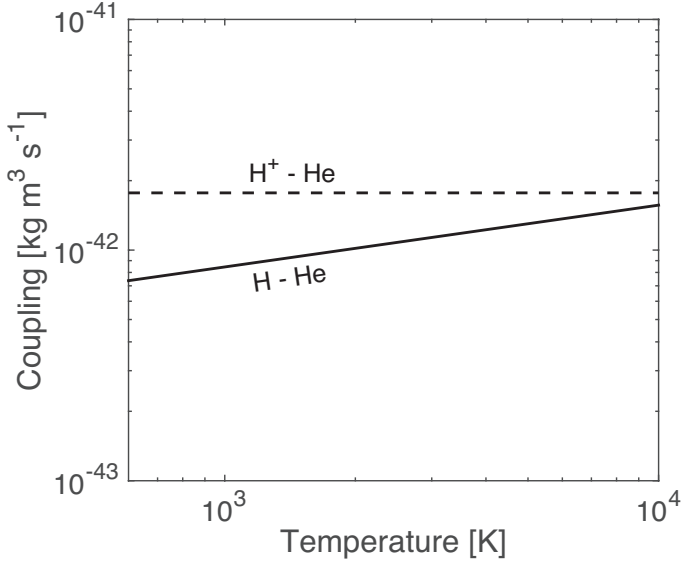}
 \caption{
Comparison between the \ce{H}-\ce{He} coupling and the \ce{H+}-\ce{He} coupling.
The \ce{H}-\ce{He} coupling (the term $kT/b$ in Eq. \ref{Mo}) is calculated from the binary diffusion coefficient \citep{Mason1970}, and the \ce{H+}-\ce{He} coupling (the term $m_{\ce{H+}}v/n_{\ce{He}}$ in Eq. \ref{Mo}) is calculated from the ion-neutral momentum transfer collision frequency \citep{Schunk1980}.
 }
 \label{Coupling}
  \end{center}
\end{figure}

The rate of momentum exchange between hydrogen and helium can be written as \citep{Schunk1980}
\begin{equation}
\frac{\delta M_{\ce{He}}}{\delta t} = n_{\ce{He}}(u_{\ce{H}}-u_{\ce{He}})n_{\ce{H}}\frac{kT}{b} + n_{\ce{He}}(u_{\ce{H+}}-u_{\ce{He}})n_{\ce{H+}}\frac{m_{\ce{H+}}v}{n_{\ce{He}}},\label{Mo}
\end{equation}
where $n$ denotes the number density, $u$ denotes the velocity, $b=1.04\times10^{18}T^{0.732}$ cm$^{-1}$ s$^{-1}$ is the binary diffusion coefficient between hydrogen and helium \citep{Mason1970}, and $v$ is the ion-neutral momentum transfer collision frequency \citep{Schunk1980}. The first term is the momentum exchange between neutral hydrogen and helium, and the second term is the momentum exchange between ionized hydrogen and helium. Helium may also be partially ionized, but the ionization fraction of helium is low up to large radius \citep[e.g.][]{Koskinen2013}. The relative efficiency between the two exchange processes is shown in Figure \ref{Coupling}. We see that ionized hydrogen or heated hydrogen can better drag helium than cold neutral hydrogen, both by up to a factor of 2. Assuming $u_{\ce{H+}}=u_{\ce{H}}$, the rate of momentum exchange can be written as
\begin{equation}
\frac{\delta M_{\ce{He}}}{\delta t} = n_{\ce{He}}(u_{\ce{H}}-u_{\ce{He}})(n_{\ce{H}}+n_{\ce{H+}})\frac{kT}{b'} ,\label{Mo1}
\end{equation}
where $b'$ is the effective binary diffusion coefficient, defined as
\begin{equation}  
\frac{kT}{b'} \equiv (1-x)\frac{kT}{b} + x\frac{m_{\ce{H+}}v}{n_{\ce{He}}}, \label{beff}
\end{equation}
where $x$ is the ionization fraction. As such, the diffusion-limited escape rate is
\begin{equation}
\phi_{\rm DL}=\frac{GM_p(m_\ce{He}-m_\ce{H})b'}{r_p^2kT},
\end{equation}
where $T$ is the temperature of the homopause. 

The fact that the efficiencies of the neutral interaction and the ion-neutral interaction are similar and only weakly depend on temperature (Figure \ref{Coupling}) allows us to choose a representative temperature and ionization fraction. As a conservative estimate for the mass fractionation effect, we use a high temperature of $10^4$ K, and an ionization fraction of 0.1, which yields $b'=8.0\times10^{20}$ cm$^{-1}$ s$^{-1}$ based on Equation (\ref{beff})

The diffusion-limited escape rate determines the difference between the two escaping components, as
\begin{equation}
\frac{\phi_\ce{He}}{X_\ce{He}} = \frac{\phi_\ce{H}}{X_\ce{H}} - \phi_{\rm DL}. \label{DiffFlux}
\end{equation}
This simple relation is derived assuming subsonic flow but is proved to be a close approximation to a transonic flow as well \citep{Zahnle1990}. This equation demonstrates that when $\phi_\ce{H} \gg X_\ce{H}\phi_{\rm DL}$, fractionation between \ce{H} and \ce{He} would be minimal, but when $\phi_\ce{H} \rightarrow X_\ce{H}\phi_{\rm DL}$, $\phi_\ce{He}\rightarrow0$, we return to the classical formula for diffusion-limited escape. Near this limit, both \ce{H} and \ce{He} escape, so the flow is not exactly diffusion-limited escape, but significant fractionation between \ce{H} and \ce{He} can still occur. With Equations (\ref{CompoFlux} and \ref{DiffFlux}) we determine the escape flux of hydrogen and helium. The solution is
\begin{eqnarray}
&& {\rm If \ \ } \Phi \le \phi_{\rm DL}X_\ce{H}m_\ce{H}4\pi r_p^2, \nonumber\\
&& \Phi_\ce{H} = \Phi, \ \ \Phi_\ce{He}=0; \label{parti1}\\
&& {\rm If \ \ } \Phi > \phi_{\rm DL}X_\ce{H}m_\ce{H}4\pi r_p^2, \nonumber\\
&& \Phi_\ce{H} = \frac{\Phi m_\ce{H}X_\ce{H} + \phi_{\rm DL} m_\ce{H}m_\ce{He}X_\ce{H}X_\ce{He}4\pi r_p^2 }{m_\ce{H}X_\ce{H}+m_\ce{He}X_\ce{He}}, \nonumber\\
&& \Phi_\ce{He} = \frac{\Phi m_\ce{He}X_\ce{He} - \phi_{\rm DL} m_\ce{H}m_\ce{He}X_\ce{H}X_\ce{He}4\pi r_p^2}{m_\ce{H}X_\ce{H}+m_\ce{He}X_\ce{He}}. \label{parti2}
\end{eqnarray}

For each time step, the mass of the atmosphere is updated by adding the escape fluxes of hydrogen and helium, and the composition and the mean molecular mass of the atmosphere are updated accordingly. Then, the new mass and mean molecular mass are used to determine the new thickness of the atmosphere and the new radius of the planet. 

\subsection{Atmosphere Chemistry Model}

Atmospheric escape and evolution can have large effects on the atmospheric chemical composition and then a planet's transmission and thermal emission spectrum, by depleting hydrogen over a long period. The fact that vertical mixing ($\sim$years) is much faster than atmospheric evolution ($\sim$billion years) allows us to treat the atmospheric chemistry calculations as ``snapshots in the evolutionary history'', with the ingredients of the atmosphere (i.e., the elemental abundances) controlled by the evolution.

We use a one-dimensional photochemistry-thermochemistry kinetic-transport model \citep{Hu2012,Hu2013,Hu2014} to explore the chemical state and spectral characteristics of a potential helium atmosphere. The photochemistry-thermochemistry kinetic-transport atmosphere model computes the atmospheric composition and the opacities self-consistently for an exoplanet, with the atmospheric elemental abundance, the efficiency of vertical mixing, and the internal heat flux as the input parameters. The model also computes the temperature profiles from the opacities using the grey-atmosphere approximation. Our atmosphere chemistry model contains two steps of computation: the first step is to compute the atmospheric composition at thermochemical equilibrium self-consistently with the temperature profile; and the second step is to use the result of the first step as the initial condition to simulate the effects of vertical mixing and photochemical processes on the atmospheric composition, with the temperature-pressure profile adjusted accordingly. After the photochemistry-thermochemistry simulation converges to a steady state, we compute synthesized spectra of the modeled exoplanet's atmospheric transmission and thermal emission with a line-by-line method. A unique feature in our photochemistry-thermochemistry model is that the model does not require specification of the main component of the atmosphere (nor the mean molecular mass) and instead the model takes the elemental abundances as the input parameters, making our model ideal for exploring the composition of hydrogen- versus helium-dominated atmospheres. 

Our general atmosphere chemistry model and its validation have been described in detail \citep{Hu2012,Hu2013,Hu2014}. To properly treat a helium-dominated atmosphere, we have made the following upgrade to the general model. We have included the \ce{H2}-\ce{He} collision-induced absorption \citep{Borysow1988} in our calculations of the opacities, and used the heat capacity of helium recommend by the NIST Chemistry Webbook (http://webbook.nist.gov/chemistry/) for calculation of the adiabatic lapse rate of helium-dominated atmospheres. We have treated helium as a chemically inert gas in our model, and included helium in calculation of the mean molecular mass.

\section{Fractionation between Hydrogen and Helium}

\begin{figure}
\begin{center}
 \includegraphics[width=0.5\textwidth]{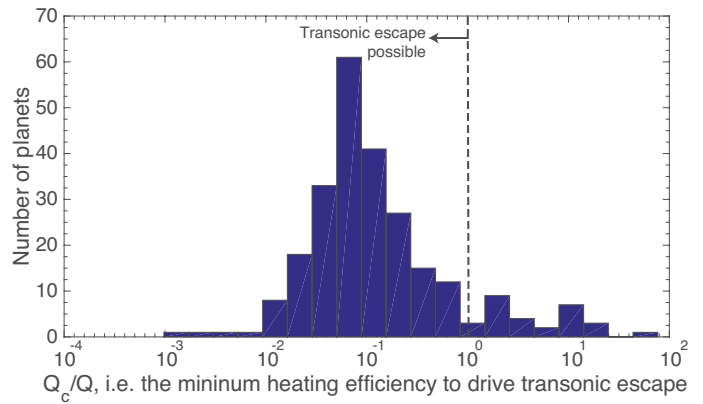}
 \caption{
The minimum heating efficiency to drive transonic hydrodynamic escape from detected exoplanets, shown as a histogram. The exoplanets having their masses and radii measured more precisely than $3-\sigma$ are counted, and the total number of planets are 254 (http://exoplanet.eu). We use Equation (\ref{Q}) to estimate the incident EUV heating rate, and assume a conservative 10-billion-year-old stellar EUV luminosity from Equation (\ref{EUV}). We use Equation (\ref{Tran_con}) to estimate the critical heating rate for transonic escape, assuming that the sonic point radius is close to the planetary radius, and that the mean molecular mass is 2 (i.e. 67\% H and 33\% He).
 }
 \label{Tran_Gen}
  \end{center}
\end{figure}

We study whether hydrodynamic escape of hydrogen can fractionate hydrogen and helium in an exoplanet atmosphere. First, most of the detected exoplanets receive enough EUV radiation to have transonic hydrodynamic escape. Figure \ref{Tran_Gen} shows that a majority of the detected exoplanets with their masses and radii measured would undergo transonic hydrodynamic escape for a moderate heating efficiency of $\sim10\%$ at an age of 10 billion years. When they are younger and the stellar EUV luminosities are higher, the planets are even more likely to experience transonic escape. This is reasonable because most of the planets have short orbital periods and locate close to their parent stars. Therefore, the transonic regime must be considered when estimating the hydrodynamic escape rate of the short-period exoplanets.

\begin{figure}
\begin{center}
 \includegraphics[width=0.5\textwidth]{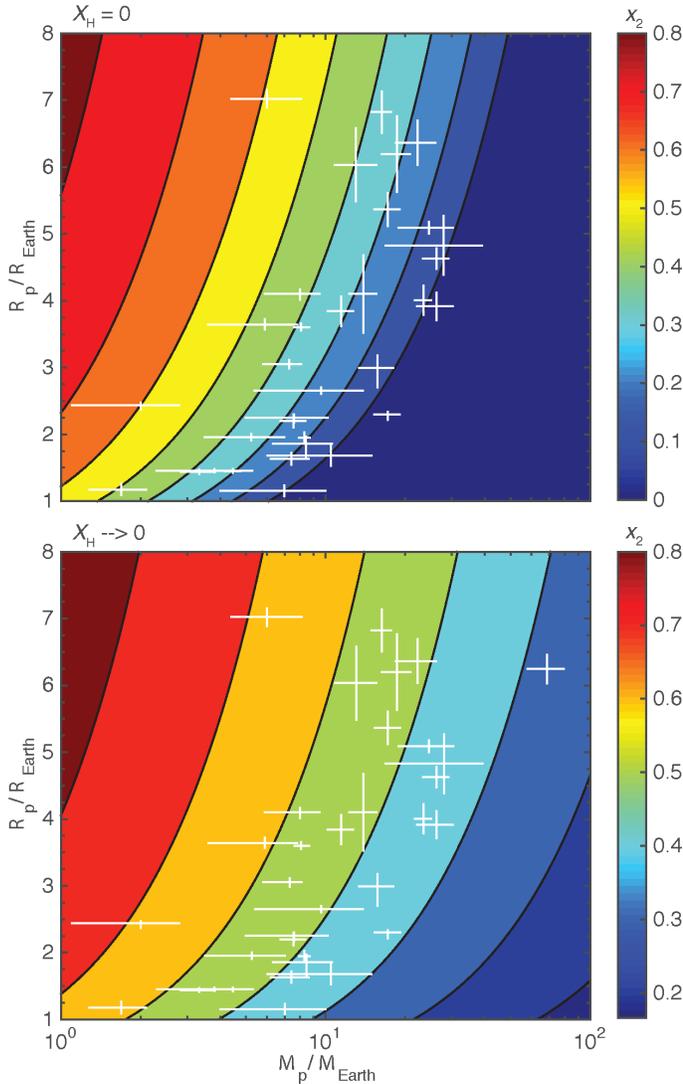}
 \caption{
Fractionation factor between hydrogen and helium for Neptune- and sub-Neptune-sized exoplanets undergoing transonic hydrodynamic escape. The color contour shows the fractionation factor ($x_2$), for which a smaller value means a more significant fractionation between hydrogen and helium. Neptune- and sub-Neptune-sized exoplanets having their masses and radii measured more precisely than $3-\sigma$ are shown.
 }
 \label{Evo_Gen}
  \end{center}
\end{figure}

Second, hydrodynamic escape from many Neptune- and sub-Neptune-sized exoplanets significantly fractionates hydrogen versus helium in their atmospheres. Assuming transonic hydrodynamic escape, we can replace $Q_{\rm net}$ by $Q_c$ in the energy-limited escape rate calculation for an estimate of the upper-limited of the escape mass flux, because increasing the stellar EUV flux would no longer increase the escape mass flux for transonic outflow \citep{Johnson2013}. Note that this upper limit is independent from the stellar irradiation, but depends on the mass and the radius of the planet, as well as the composition of the atmosphere. We then partition this mass flux to the escape fluxes of hydrogen and helium, using Equations (\ref{parti1} and \ref{parti2}), and calculate the fractionation factor between hydrogen and helium, $x_2$, which is defined as
\begin{equation}
x_2 = \frac{\phi_{\rm He}/\phi_{\rm H}}{X_{\rm He}/X_{\rm H}}.
\end{equation}
$x_2$ depends on the instantaneous volume mixing ratio between hydrogen and helium, and takes a value between 0 and 1. $x_2=1$ means no fractionation, and $x_2=0$ means the highest degree of fractionation (i.e., no helium escaping with hydrogen). We show the results in Figure \ref{Evo_Gen}.

Most of the Neptune- and sub-Neptune-sized planets have a small fractionation factor between hydrogen and helium. This means that hydrodynamic escape from their atmospheres would deplete hydrogen and enrich helium. Since the assumed escape rate is an upper limit, we are conservative in estimating this fractionation effect. For a smaller escape rate, the fractionation effect would be even more significant. We also note that the fractionation effect becomes milder as the hydrogen mixing ratio decreases. However, the asymptotic value of $x_2$ is less than unity (Figure \ref{Evo_Gen}). This implies that with unlimited time and no hydrogen source, the process can in principle remove hydrogen in the atmosphere to an arbitrarily low abundance. 

Therefore, fractionation between hydrogen and helium is significant for hydrodynamic escape from many exoplanets. This does not imply these planets have helium atmospheres, however, as additional conditions have to be met. Our atmosphere evolution model serves to explore the conditions to form helium atmospheres for specific planets. Neptune- and sub-Neptune-sized planets can have small hydrogen and helium envelopes, and the fractionation factors for them are substantially smaller than unity. Note that the fractionation factors for them are not zero, so they are technically not in the regime of diffusion-limited escape. Rather, hydrodynamic escape from these planets is highly affected by the diffusion separation between hydrogen and helium. We suggest that fractionation between hydrogen and helium is an important process for hydrogen-helium atmospheres of short-period exoplanets undergoing hydrodynamic escape. The effect of this fractionation, as later discussed for the cases of GJ~436~b, depends on the initial mass of the atmosphere, and is likely to be significant for Neptune-sized and smaller exoplanets.

\section{A Helium Atmosphere Model for GJ~436~b}

The observational characterization of atmospheric compositions of Neptune- and sub-Neptune-sized exoplanets is ongoing \citep[e.g.][]{seager2010b}, and provides the opportunity to search for highly evolved exoplanet atmospheres and test our hypothesis of the formation of helium atmospheres. One of the most observed Neptune-sized exoplanets is GJ~436~b \citep{Gillon2007,Deming2007,Demory2007,Alonso2008,Pont2009,Caceres2009,Ballard2010,Stevenson2010,Beaulieu2011,Knutson2011,Knutson2014,Lanotte2014,Morello2015}. This planet orbits a nearby M dwarf star and has a mass of 23.2 $M_{\oplus}$ and a radius of 4.22 $R_{\oplus}$ \citep{Torres2008}. The bulk density of the planet implies that the planet should have a thick atmosphere or envelope made of hydrogen and/or helium, which itself has a mass of at least $10^{-4}\sim10^{-3}$ of the planet's mass \citep{Nettelmann2010}.

\subsection{Puzzling Emission Features of GJ~436~b}

The composition of GJ~436~b's atmosphere has been a puzzle. A broad-band emission spectrum of the planet in 3-30 $\mu$m was obtained from {\it Spitzer} \citep{Stevenson2010, Lanotte2014}, which indicated that the planet's atmosphere is poor in \ce{CH4} and rich in \ce{CO} \citep{Madhusudhan2011,Line2014}. The key spectral feature is a detected emission flux in the 3.6 $\mu$m band (\ce{CH4} absorption band) and non-detection in the 4.5 $\mu$m band (\ce{CO} absorption band). We note that the reported 3.6 $\mu$m emission flux of \cite{Lanotte2014} was significantly lower than that of \citet{Stevenson2010}, though both values were derived from the same observation (see Appendix \ref{Observation} for discussion). In addition, a recent transmission spectrum of the planet in 1-2 $\mu$m was obtained from {\it HST}, and the spectrum lacks \ce{H2O} or \ce{CH4} features \citep{Knutson2014}. These observations and their interpretation lead to a theoretical challenge: even when accounting for disequilibrium processes \citep{Line2011}, atmospheric chemistry models uniformly predict most carbon should be in the form of \ce{CH4}--not \ce{CO}--in a solar-abundance hydrogen and helium background atmosphere. 

A solar-abundance hydrogen and helium atmosphere on GJ~436~b cannot produce a spectrum consistent with either dataset of the spectra \citep{Stevenson2010, Lanotte2014}, because the chemical equilibrium models would predict most of carbon to be in the form of \ce{CH4}, but the observations indicate otherwise. Including disequilibrium processes like strong vertical transport and efficient photochemical reactions, or exploration of temperature variation in reasonable ranges, cannot address the ``missing-methane'' problem \citep{Madhusudhan2011,Line2011,Agundez2014}. To fit the emission spectrum of \cite{Stevenson2010}, the mixing ratio of \ce{CH4} between 1 and 0.01 bar would need to below $\sim1$ ppm based on retrieval studies allowing the compositions and the temperature profiles to vary freely and independently \citep{Madhusudhan2011, Moses2013a}. The retrieval results of fitting to the more recent dataset of \cite{Lanotte2014} are not available, but one could imagine that more \ce{CH4} could be allowed in the atmosphere because of the smaller 3.6-$\mu$m eclipse depth.

\cite{Moses2013a} presented a series of models for the atmospheric compositions and emission spectra of GJ~436~b, using the PHOENIX atmospheric model \citep{Barman2005} for the temperature-pressure profiles. To obtain a mixing ratio of methane less than 1 ppm, a metallicity 10,000 times greater than the solar atmosphere would be required \citep[][Figure 12]{Moses2013a}. Such a 10,000x solar atmosphere would be \ce{CO2}-, \ce{CO}-, \ce{N2}- and \ce{H2O}-dominated, and would have a very high mean molecular mass. 

However, this hypothetical world cannot be GJ~436~b, for such an extravagant metallicity violates the constraints on the bulk density of the planet in question. The radius of an equally massive planet with a \ce{CO}-, \ce{CO2}- and \ce{H2O}-dominated atmosphere would be much smaller than the measured radius of GJ~436~b. Detailed models of the interior structures of Neptune-sized exoplanets have also ruled out such a dense outer envelope \citep{Rogers2010, Nettelmann2010}. In particular, \cite{Nettelmann2010} calculated a high-temperature endmember scenario, in which the temperature profile was assumed to be adiabatic and the temperature at 1 bar was assumed to be 1300 K (series G). This scenario had a temperature as high as 5500 K at 1000 bar, but the minimum mass of hydrogen and helium for this model was still at least 0.01\% of the planetary mass. A 10,000x solar outer envelope would not be permitted even when a fairly high internal temperature is considered. 

In addition to raising the metallicities, \cite{Moses2013a} then assumed an {\it ad hoc} temperature profile, and proposed that an \ce{H2}- and \ce{He}-dominated atmosphere 300 times more metal-rich than the solar atmosphere could be consistent with the emission spectrum reported by \cite{Stevenson2010}. To produce a sufficiently low mixing ratio of \ce{CH4}, the temperature of such an atmosphere at 0.1 bar would have to be $\sim1300$ K, greater than radiative-convective model predictions by at least 200 K. In other words, the 300x solar model of \cite{Moses2013a} likely assumed an unphysically large temperature at 0.1 bar. The tidal heat due to the planet's orbital eccentricity (e=0.16) is deposited at a much deeper pressure and cannot raise the temperature at 0.1 bar \citep{Agundez2014}. When using reasonable temperature profiles, raising the metallicities to the extent still allowed by the planet's bulk density cannot reduce the mixing ratio of \ce{CH4} in the observable part of the atmosphere to less than 1 ppm. 

We also note that some of the self-consistent models for 300x and 1000x solar abundances in \cite{Moses2013a}, without assuming an {\it ad hoc} temperature profile, appear to be consistent with the more recent dataset of the emission spectrum \citep{Lanotte2014}. 

Alternatively, one could explain GJ~436~b's emission spectrum by the aforementioned scenario of atmospheric evolution culminating in a helium-dominated atmosphere. Once such a state is achieved, the scarcity of atmospheric hydrogen will cause the main molecular carrier of carbon to be \ce{CO} rather than \ce{CH4}, explaining the emission spectrum.

\subsection{Chemistry and Spectrum}

Using the atmosphere chemistry model, we explore a wide range of parameters that define the atmospheric composition of GJ~436~b, including the hydrogen elemental abundance of the atmosphere (by number, denoted as $X_\ce{H}$), the metallicity of the atmosphere (denoted as $X_\ce{M}$), the carbon to oxygen ratio of the atmosphere (denoted as $X_\ce{C}/X_\ce{O}$), the intrinsic temperature (to specify the internal heat flux, denoted as $T_{\rm int}$), and the eddy diffusivity (to specify the efficiency of vertical mixing, denoted as $K_{\rm zz}$). In this work we consider only H, He, C, and O species, so that $X_\ce{He}=1-X_\ce{H}-X_\ce{M}$ and $X_\ce{M}=X_\ce{C}+X_\ce{O}$. Nitrogen, sulfur, and other metal species may also contribute to the spectrum. Our selection suffices because the key observational feature to explain is lack of methane absorption. For each atmosphere scenario that we consider, we use the thermochemistry-photochemistry model to compute the steady-state molecular composition from $10^3$ bar to $10^{-8}$ bar. We verify that the lower boundary at $10^3$ bar is sufficient to maintain thermochemical equilibrium at the lowest atmosphere layer for each simulation. For the stellar input spectrum, we use the latest {\it HST} measurement of the UV spectrum of GJ 436 \citep{France2013} and the NextGen simulated visible-wavelength spectrum \citep{Allard1997} of an M star having parameters closest to those of GJ 436 (i.e., effective temperature of 3600 K, surface gravity log($g$)=5.0, and solar metallicity). 

\begin{table*}
\footnotesize
\centering
\caption{The parameters and the modeled compositions for the atmospheric models shown in Figure \ref{Model}.}
\begin{tabular*}{\hsize}{@{\extracolsep{\fill}}cccc|cccccc|c|c}
\hline\hline
\multicolumn{4}{l|}{Parameter} & \multicolumn{6}{l|}{Molecular Composition at 0.01-1 bar} & $\chi^2/n$ & Color \\
\hline
$X_\ce{H}$ & $X_\ce{M}$ & $X_\ce{C}/X_\ce{O}$ & $T_{\rm int}$ (K) & \ce{He} & \ce{H2} & \ce{H2O} & \ce{CH4} & \ce{CO} & \ce{CO2} & &\\
\hline
$1\times10^{-4}$ & $1\times10^{-2}$ & 0.95 & 60 & 0.9951 & $4.8\times10^{-5}$ & $2.5\times10^{-6}$ & $3.1\times10^{-10}$ & $4.6\times10^{-3}$ & $2.5\times10^{-4}$ & 2.5\tablenote{Comparison to the data of \cite{Stevenson2010}} & r \\
$3\times10^{-5}$ & $1\times10^{-3}$ & 0.90 & 20 & 0.9995 & $1.4\times10^{-5}$ & $1.3\times10^{-6}$ & $8.8\times10^{-9}$ & $4.2\times10^{-4}$ & $5.1\times10^{-5}$ & 7.1$^{\rm a}$ & m \\
$1\times10^{-3}$ & $1\times10^{-3}$ & 0.9997 & 60 & 0.9990 & $4.8\times10^{-4}$ & $2.8\times10^{-6}$ & $6.5\times10^{-6}$ & $4.9\times10^{-4}$ & $3.9\times10^{-6}$ & 2.7\tablenote{Comparison to the data of \cite{Lanotte2014}} & b \\
$3\times10^{-1}$ & $1\times10^{-1}$ & 0.70 & 60 & 0.76 & 0.17 & $1.8\times10^{-2}$ & $4.1\times10^{-4}$ & $4.7\times10^{-2}$ & $5.1\times10^{-3}$ & 3.1$^{\rm b}$ & g \\
\hline\hline
\end{tabular*}
\label{Para}
\end{table*}

\begin{figure}
\begin{center}
 \includegraphics[width=0.5\textwidth]{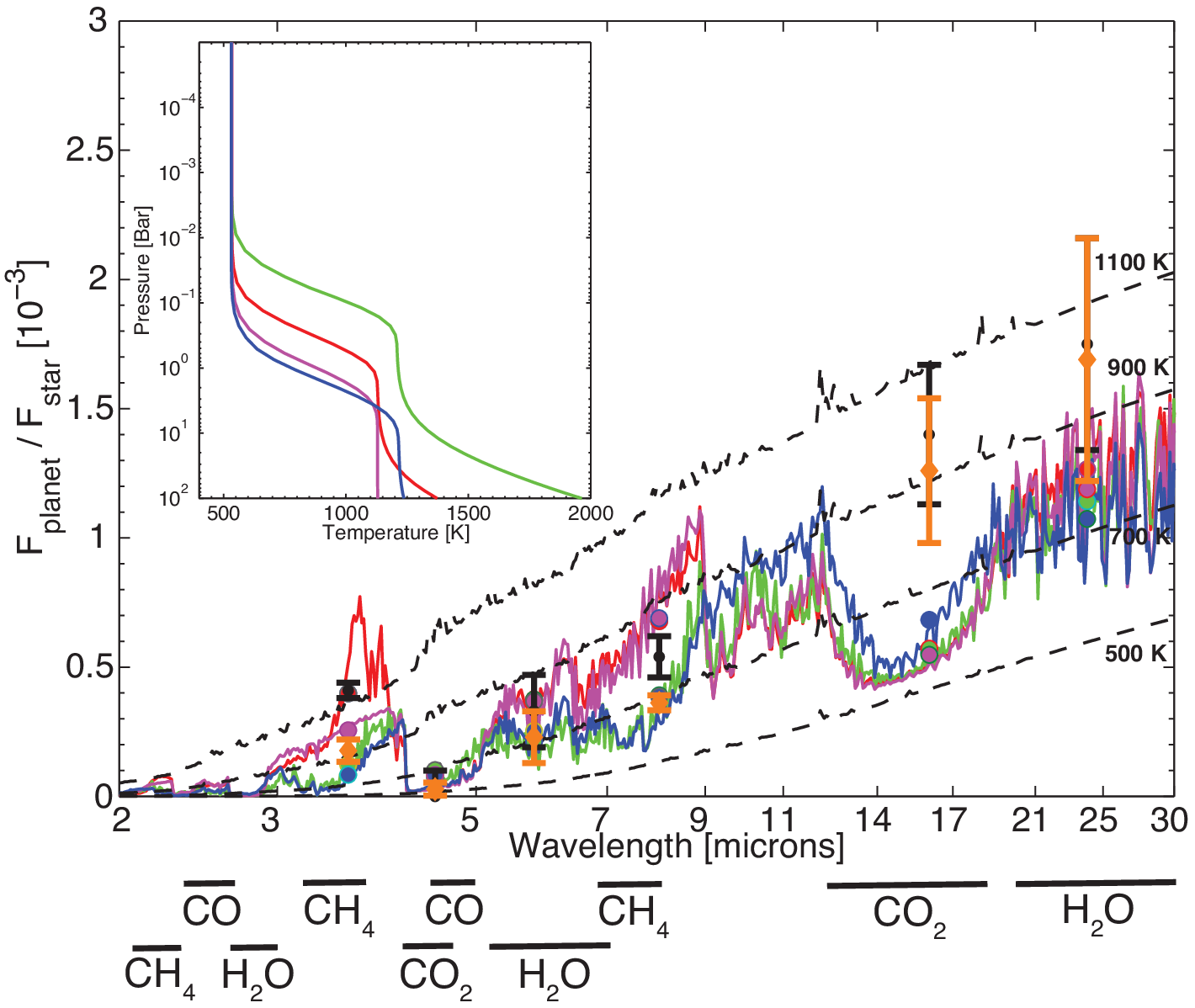}
 \caption{
Comparison of model spectra of the dayside emission of GJ~436~b with observations. 
The reported values for the eclipse depths are shown in black \citep{Stevenson2010} and orange \citep{Lanotte2014}.
The filled circles are the model-predicted occultation depths with the {\it Spitzer} bandpass incorporated. The wavelength ranges of major molecular absorption bands are shown below the horizontal axis.
The spectra are based on atmosphere models with parameters tabulated in Table \ref{Para}, and the corresponding temperature-pressure profiles are shown in the inserted panel. The thin dashed lines show the eclipse depths if the planet emits as a black body with temperatures of 500-1100 K.
The red line shows the best-fit model to the data of \cite{Stevenson2010}, which has $X_\ce{H}=10^{-4}$ and an internal heat flux. The magenta line shows a model with the same composition but without the internal heat flux.
The blue line shows the best-fit model to the data of \cite{Lanotte2014}, and the green line shows the a model of \ce{H2}-rich metal-rich atmospheres ($X_\ce{H}\ge0.3$, $X_\ce{M}\ge0.1$).
The thermal emission measurement at the 3.6 $\mu$m band constrains on the potential planetary scenarios: the value reported by \cite{Stevenson2010} suggests a planet having an \ce{H2}-poor atmosphere and internal heating; but the value reported by \cite{Lanotte2014} allows both the scenarios of an \ce{H2}-poor atmosphere and the scenarios of an \ce{H2}-rich metal-rich atmosphere.}
 \label{Model}
  \end{center}
\end{figure}

\begin{figure}
\begin{center}
 \includegraphics[width=0.5\textwidth]{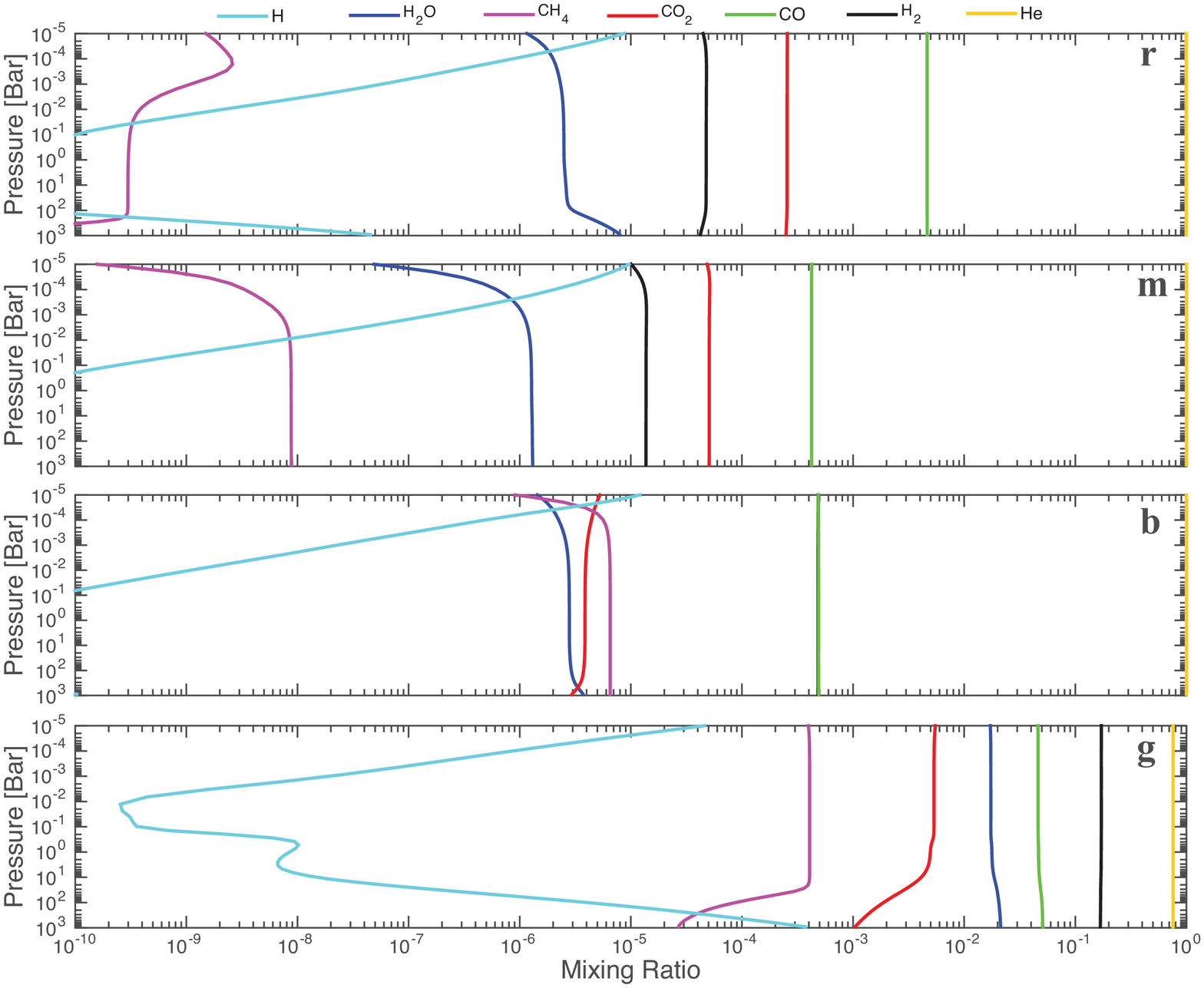}
 \caption{
Modeled compositions of a helium atmosphere on GJ~436~b. The mixing ratios of important species are shown as a function of pressure. Each color corresponds to one species, and each panel corresponds to a model shown in Figure \ref{Model} and summarized in Table \ref{Para}. The panel label indicates the color of the synthesized spectrum in Figure \ref{Model}. In the third panel (``b''), \ce{H2} and \ce{CO} lines overlap because their mixing ratios are very close.
}
 \label{ModelChem}
  \end{center}
\end{figure}

\begin{figure}
\begin{center}
\includegraphics[width=0.5\textwidth]{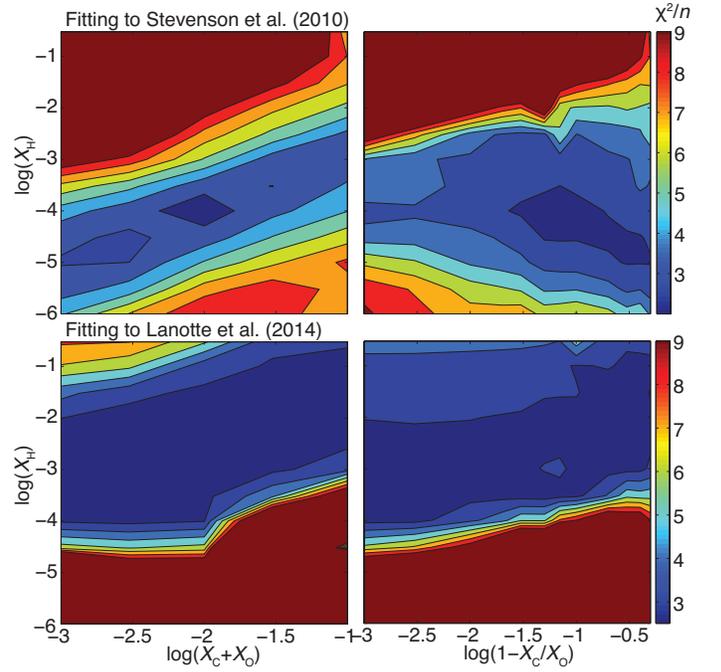}
 \caption{
Constraints on the composition of the atmosphere on GJ~436~b.
The color contours show the goodness of fit by comparing atmospheric models to the dayside emission \citep{Stevenson2010,Lanotte2014}. 
The goodness of fit is defined as the $\chi^2/n$, where $n=6$ is the number of observations. The axes define the mixing ratios of hydrogen, carbon, and oxygen with respect to the total number of atoms, and the mixing ratio of helium is $X_\ce{He}=1-X_\ce{H}-X_\ce{C}-X_\ce{O}$. The axis for the carbon to oxygen ratio is shown in terms of $\log(1-X_\ce{C}/X_\ce{O})$ to emphasize the sensitivity when the carbon to oxygen ratio approaches 1.
The upper two panels compare with the reported values of \cite{Stevenson2010} and the bottom two panels compare with the reported values of \cite{Lanotte2014}.
The parameters not shown are marginalized, including the eddy diffusion coefficient ranging from $10^6$ to $10^9$ cm$^2$ s$^{1}$, which are reasonable values for deep atmospheres according to the free-convection and mixing-length theories \citep{Gierasch1985}.
Both datasets allow \ce{H2}-poor scenarios, but the data of \cite{Lanotte2014} also allow \ce{H2}-rich metal-rich scenarios.
 }
 \label{Constraint}
  \end{center}
\end{figure}

\begin{figure}
\begin{center}
 \includegraphics[width=0.5\textwidth]{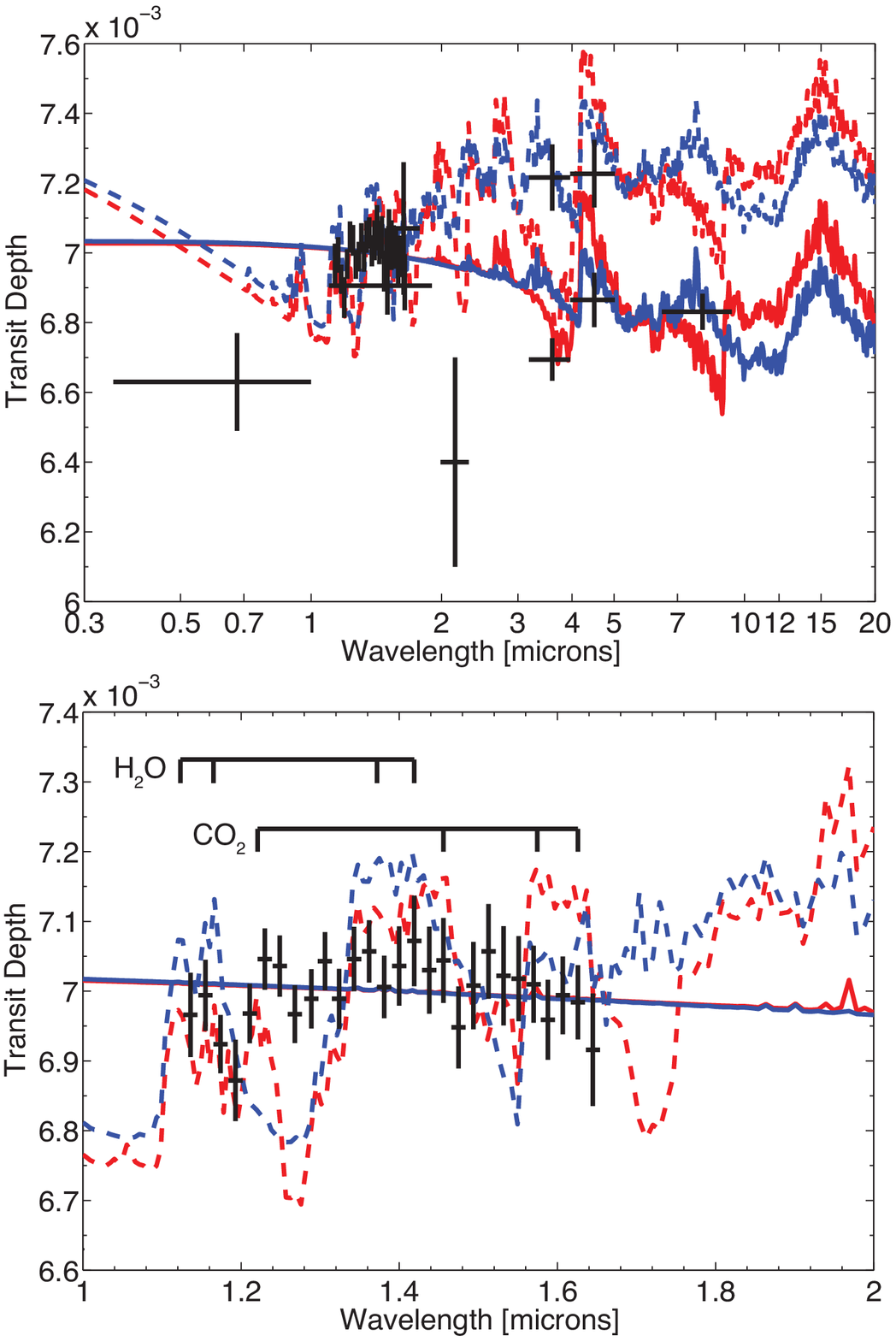}
 \caption{
Comparison of model transmission spectra of GJ 436 b with observations \citep{Ballard2010,Pont2009,Alonso2008,Caceres2009,Knutson2011,Knutson2014}. The upper panel shows the transmission spectra from 0.3 to 20 $\mu$m and the lower panel provides a zoom-in view for the wavelength of 1-2 $\mu$m featuring recent {\it HST} observations \citep{Knutson2014}.
The model spectra are based on self-consistent atmosphere models with parameters tabulated in Table \ref{Para}, i.e., the red lines corresponding to the best-fit model to \cite{Stevenson2010}, and the blues lines corresponding to the best-fit model to \cite{Lanotte2014}.
The dashed lines show the transmission spectra of a clear atmosphere, dominated by molecular absorption features.
The solid lines show the transmission spectra of an atmosphere that has an aerosol layer at pressures between 1 and 100 mbar, with an aerosol opacity of 0.001 cm$^2$ g$^{-1}$ at 3 $\mu$m. 
The aerosol optical property is computed for a material with a refractive index of 1.001, and a lognormal size distribution with a mean diameter of 0.1 $\mu$m and a size dispersion factor of 2, using the method detailed in \cite{Hu2013}.
The vertical optical depth produced by such an aerosol layer is maximally 0.1 for wavelengths longer than 3 $\mu$m, yielding no impact on the interpretation of the thermal emission spectrum.
A helium-dominated atmosphere of GJ~436~b must have an aerosol layer to be consistent with the featureless transmission spectrum at 1-2 $\mu$m. 
 }
 \label{Transmission}
  \end{center}
\end{figure}

We find that a helium-dominated atmosphere with a low hydrogen abundance and close-to-solar carbon and oxygen abundances can adequately fit both datasets of GJ~436~b's emission spectrum \citep{Stevenson2010, Lanotte2014}. At the same time a very different model also generally fits the more recent dataset \citep{Lanotte2014}, a hydrogen-rich atmosphere with highly super-solar carbon and oxygen abundances (Figure \ref{Model} and Figure \ref{ModelChem}). There is a strong residual discrepancy between the data and both models at 16 $\mu$m, and it can be further reduced if the temperature profile is allowed to adjust freely (Appendix \ref{Residual}).

Both the helium atmosphere (i.e., the \ce{H2}-poor model) and the \ce{H2}-rich metal-rich model ensure the predominance of \ce{CO} over \ce{CH4} as the main form of carbon, the key indication of the planet's emission spectrum. Our chemical models indicate that the atmosphere would be depleted in methane when $X_\ce{H}<X_\ce{M}\sim X_\ce{C}+X_\ce{O}$ (Figure \ref{Constraint}). Under this condition, methane would have to compete with water for the limited resource of hydrogen. This condition generalizes the result of a previous atmospheric chemistry model of this planet \citep{Moses2013a}; instead of increasing the metallicity $X_\ce{M}$, we suggest decreasing $X_\ce{H}$ could also make this condition satisfied. In addition, the weak absorption features at the 5.8-$\mu$m and 8.0-$\mu$m bands seen in both datasets suggest that the atmosphere should still have \ce{H2O} and \ce{CH4}, and lead to a lower bound of $X_\ce{H}$ (Figure \ref{Constraint}).

Both the \ce{He} atmosphere and the \ce{H2}-rich atmosphere scenarios require an aerosol layer at $\sim1$ mbar pressure to generate the observed featureless transmission spectrum (Figure \ref{Transmission}). This is because a hydrogen and/or helium atmosphere, with a mean molecular mass of 2 - 4, would produce large absorption features in transmission if it does not have aerosols. We find that a thin aerosol layer having a vertical optical depth less than 0.1 at mid-infrared wavelengths could produce the featureless transmission spectrum observed at 1-2 $\mu$m, and also consistent with the broadband transit depths at 3.6, 4.5, and 8.0 $\mu$m (Figure \ref{Transmission}). Such an aerosol layer would have little impact to the emission spectra. A thicker aerosol layer is possible and consistent with the transmission spectrum. We have assumed that the atmosphere to be free of thick clouds in the calculation of thermal emission, because the atmosphere would be too hot to form water clouds and too cold to form silicate clouds \citep{Sudarsky2003}. Even if thick clouds were present, the detected emission flux in the 3.6 $\mu$m band and non-detection in the 4.5 $\mu$m band still require the main molecular carrier of carbon to be \ce{CO} and not \ce{CH4}.

The large emission flux at 3.6 $\mu$m reported by \cite{Stevenson2010} corresponds to a brightness temperature greater than 1100 K. With our model atmosphere, this thermal emission emerges from the deep atmosphere (Figure \ref{Model}), which appears to be only possible for a clear helium-dominated atmosphere highly depleted in hydrogen and having close-to-solar abundances of carbon and oxygen. For such an atmosphere, the mixing ratio of methane is extremely low, and there is essentially little opacity at 3.6 $\mu$m, thereby creating an infrared window into the deep atmosphere. An internal heat flux $\sim10$ times Earth's geothermal flux may thus drive a deep adiabatic temperature profile and lead to a high emission flux in this window region. In our best-fit model, the emission of a clear helium-dominated atmosphere having a very low amount of methane can show high brightness temperatures between 3.6 and 4.0 $\mu$m, ranging from 1100 to 1400 K (Figure \ref{Model}). Comparing with the temperature-pressure profile, this corresponds to emitting from as deep as 1-100 bars at these wavelengths. Probing such deep atmospheres is probably one of the unique features of the proposed helium-dominated atmosphere. For this reason, a small energy flux from the interior that drives a deep adiabatic temperature profile could lead to appreciable changes at specific wavelengths of the emerging infrared spectrum.

We note that the grey-atmosphere approximation does not capture this infrared window, which makes our calculation of the temperature profile no longer consistent with the atmospheric composition. As a plausibility check, we calculate the total infrared emission flux from the planet, with and without the spikes at 3.6-4.0 $\mu$m. We find the case with a deep adiabat emits 6.6\% more energy through this window than the case without a deep adiabat. This additional emission energy, if solely attributed to the internal heat flux, would correspond to an intrinsic temperature of 370 K, much higher than the 60 K assumed in the grey-atmosphere approximation. A non-grey radiative-convective model is required to accurately calculate the temperature profile and the internal heat flux indicated by the emission feature.

One might ask whether probing such deep atmosphere is ever possible. We show that this is possible for a helium-dominated and hydrogen-poor atmosphere in terms of molecular absorption, because the common opacity sources at this wavelength range, including \ce{CH4} and \ce{H2}-\ce{H2} collision-induced absorption, disappear in the helium-dominated atmosphere. \ce{H2}-\ce{He} collision-induced absorption is insignificant. The lack of absorbers thus provides an infrared window. One might also wonder, given that the model has a 10x solar metallicity, whether clouds could form above $\sim100$ bar and block this emission window \citep{Burrows1999}. Comparing the modeled temperature profiles to the condensation temperatures of the materials commonly suggested to form clouds in the atmospheres of irradiated exoplanets, including \ce{H2O}, \ce{Mg2SiO4}, \ce{MgSiO3}, and \ce{Fe}, we find that GJ~436~b is too hot to have water clouds, and too cold to have silicate or iron clouds above 100 bars \citep{Sudarsky2003}. A ``cloud deck'', if existing, is then likely to be deeper than what we are concerned with. A cautionary note is that strong tidal heating may raise the temperature of the deep convective layer \citep{Agundez2014}, and raise the altitude of this potential cloud deck, causing the clouds to affect the infrared spectra.

\subsection{Formation of a Helium Atmosphere}

\subsubsection{Current Escape}

GJ~436~b is undergoing transonic hydrodynamic escape. The total EUV flux received by GJ~436~b is $7.3\times10^{15}$ W estimated from Equations (\ref{EUV}), and this estimate is fully consistent with X-ray and Ly-$\alpha$ flux measurements of GJ~436 (Appendix \ref{Xray}). Assuming $\gamma=5/3$ for an atomic flow, $c_c\sigma_c=5\times10^{-20}$ m$^2$, $Kn_m\sim1$ (valid when the heat is primarily absorbed over a broad range of radius below $r_*$), and $r_*\sim r_0$, we estimate the critical heating rate to be $Q_c=6.5\times10^{14}$ W for a helium-dominated atmosphere, and $2.3\times10^{14}$ W for a hydrogen-dominated atmosphere from Equation (\ref{Tran_con}). We find that as long as $\eta>0.09$, the heating rate would be greater than the critical rate and the current hydrodynamic escape would be transonic. Since the earlier EUV fluxes have been greater than the current flux, hydrodynamic escape on GJ~436~b would have always been transonic.

Our calculation of the escape rate is consistent with the recent {\it HST} observation of the GJ~436 system in the Lyman-$\alpha$ absorption \citep{Kulow2014}. The escape rate of \ce{H} derived from the Ly-$\alpha$ transit light curve is $4\times10^6\sim2\times10^7$ g s$^{-1}$, assuming an ionization fraction of 0.1. In comparison, the standard energy-limited escape rate of a hydrogen-dominated atmosphere calculated from Equation (\ref{TotalFlux}) would be $10^8\sim10^{10}$ g s$^{-1}$ for an efficiency ranging from 0.01 to 1. Even for a very low efficiency, the standard formula overestimates the escape rate. In our model, the current escape rate is $7.3\times10^4\sim7.3\times10^6$ g s$^{-1}$ for \ce{H} and $6.8\times10^8\sim6.9\times10^8$ g s$^{-1}$ for \ce{He}, for $X_\ce{H}$ ranging between $10^{-4}$ and $10^{-2}$, consistent with the observation. These calculations assume a heating efficiency of 10\%, but varying the efficiency from a few \% to 50\% does not change the results significantly, because the escape rate is limited by $Q_c$ for the transonic flow. Here, the apparent agreement is only a proof of concept rather than observational confirmation, because the reported absorption depth is measured after the optical transit, and stellar variabilities could significantly affect the planetary signal \citep{Loyd2014,Llama2015}. Also, the reported high velocity of the escaping hydrogen corona (60-120 km s$^{-1}$) does not indicate the outflow to be supersonic by itself. The absorption at high Doppler shifts may very well be due to absorption in the wings of the line profile \citep{BenJaffel2010,Koskinen2010}, or produced by charge exchange with stellar wind protons \citep{Holmstrom2008,Kislyakova2014}.

\subsubsection{Formation Conditions}

\begin{figure}
\begin{center}
\includegraphics[width=0.5\textwidth]{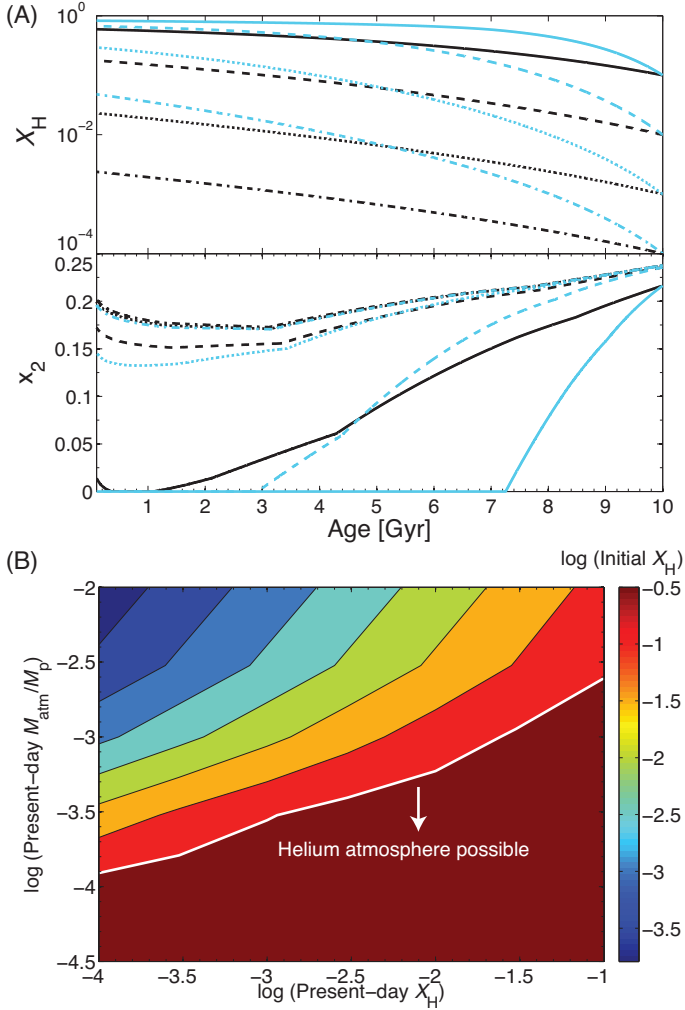}
 \caption{
Fractionation between helium and hydrogen via hydrodynamic escape of GJ~436~b, as a function of the present-day $X_\ce{H}$ and the present-day mass of the atmosphere. Panel (A) shows the history of the hydrogen abundance of the atmosphere and the fractionation ratio between helium and hydrogen by escape ($x_2$), for an atmosphere presently having a mass of $10^{-3}$ (black lines) or $3\times10^{-4}$ (blue lines) planetary mass and various present-day $X_\ce{H}$ distinguished by line types. The total mass loss is approximately $10^{-3}$ planetary mass in these models. Panel (B) shows the calculated initial $X_\ce{H}$ by color contours, assuming an escape efficiency of 10\% and an age of 10 billion years. We have assumed the atmosphere to be isothermal with a temperature of 2000 K. Even a temperature of 1000 K yields only a very minor alteration, indicating the isothermal assumption is adequate. We caution that the interior and the atmosphere may cool down and contract significantly during the first billion years, and our calculations may not capture this early phase of evolution. When the mass of the atmosphere is $10^{-3}$ planetary mass or less, hydrodynamic escape can reduce the hydrogen abundance of the atmosphere by orders of magnitude.
  }
 \label{Evolution}
  \end{center}
\end{figure}

We find that the fractionation factor between helium and hydrogen ($x_2$) is $\sim0.2$ at present for a hydrogen abundance ranging from 0.1 to $10^{-4}$, and has remained below $\sim0.2$ for the entire evolution history (Figure \ref{Evolution}). The fractionation factor is significantly smaller than unity, meaning that hydrodynamic escape disproportionally remove hydrogen from the atmosphere.

The cumulative effect of the selective loss of hydrogen via hydrodynamic escape depends on the mass of the atmosphere. If the initial mass of the atmosphere is $10^{-3}$ planetary mass or less, hydrodynamic escape would decrease the hydrogen abundance of the atmosphere by more than one order of magnitude on a time scale of a few billion years (Figure \ref{Evolution}). The change of the hydrogen abundance is greater for a smaller initial mass. Therefore, the timescale required to significantly reduce the hydrogen abundance from around the cosmic value to $10^{-2}\sim10^{-4}$ by hydrodynamic escape is sensitive to the mass of the atmosphere. For example, hydrodynamic escape for 10 billion years can fractionate an atmosphere that is $10^{-3}$ planetary mass or less (Figure \ref{Evolution}). Similarly, hydrodynamic escape for 5 billion years can fractionate an atmosphere that is $3\times10^{-4}$ planetary mass. 

The mass fractionation can be more efficient if the homopause is well below the EUV absorption altitude and has a temperature close to the planet's equilibrium temperature. For a temperature of 600 K, additional calculations indicate a fractionation factor between helium and hydrogen of $\sim0.1$. In that case, hydrodynamic escape for 10 billion years can well fractionate an atmosphere that is $3\times10^{-3}$ planetary mass.

The amount of hydrogen and helium accreted by the planet when it formed depends the disk dissipation time, the orbital location, and the mass of the planet's core \citep{Rafikov2006,Hansen2012,Ikoma2012,Inamdar2014}. If the planet formed at large orbital separation and migrated to close-in orbits, \citet{Rafikov2006} shows that a 20-$M_{\oplus}$ core would only accrete $\sim0.1$ $M_{\oplus}$ of hydrogen and helium at 10 AU if the core has a high luminosity and the accretion is fast. This makes the initial mass of the hydrogen-helium envelope 0.5\% the planetary mass. Engulfing planetesimals or radiogenic heating would increase the core luminosity and reduce the amount of hydrogen and helium accreted. After accretion, the initial atmosphere may also be reduced in mass by photoevaporation \citep[e.g.][]{Lopez2012,Fortney2013}. If the planet formed in situ \citep{Hansen2012}, the amount of hydrogen and helium accreted would be on the orders of $10^{-3}\sim10^{-2}$ the core mass, and this initial envelope is typically reduced in mass by giant impacts by $1\sim2$ orders of magnitudes \citep{Inamdar2014}. Both formation scenarios could provide conditions to form the helium atmosphere.

It is also important to check how long the proposed helium-dominated atmosphere would be stable against further transonic hydrodynamic escape. Apparently the total mass loss for 10 billion years is on the order of $10^{-3}$ planetary mass. Therefore, it is conceivable that a helium-dominated envelope of $10^{-3}$ planetary mass would be stable for billions of years. Figure \ref{Evolution} also shows evolution scenarios in which the planet has already had a helium-dominated atmosphere $\sim5$ billion years ago, and the helium-dominated atmosphere persists to the present day. Therefore, although hydrodynamic escape might eventually deplete helium from the planet, the evolutionary phase when helium dominates the outer envelope can be fairly long.

We therefore propose that exoplanet GJ~436~b may have a helium-dominated atmosphere that evolved from a primordial hydrogen and helium envelope. The proximity of the planet to its parent star provides the conditions to maintain transonic hydrodynamic escape throughout the evolution history. The mass and the size of the planet determine whether the escape rate has been close to the diffusion-limited escape rate of hydrogen. As a result, the planet has experienced disproportional loss of its primordial hydrogen. Some of the primordial helium has also been lost during this evolution, but some may have remained.

\section{Discussion: Detecting Evolved Exoplanet Atmospheres via Transmission}

The planetary evolution scenario for GJ~436~b can be a general process happening on many Neptune and sub-Neptune-sized exoplanets. Due to depletion of hydrogen, the abundances of major carbon- and oxygen-bearing species would change by orders of magnitude \citep{Hu2014}. For example, the most important trace gases in an \ce{H2}-dominated atmosphere are \ce{CH4} and \ce{H2O} at equilibrium temperatures lower than $\sim1000$ K (defined for zero albedo and full heat redistribution). When the planet loses hydrogen, the dominant carbon- and oxygen-bearing species change to \ce{CO}, high-order hydrocarbons, \ce{CO2}, and \ce{O2}. 

With the recognition that primordial helium may persist as a planet loses hydrogen, we propose that the highly-evolved, helium-dominated atmospheres can be detected by measuring the transmission spectra of Neptune- and sub-Neptune-sized exoplanets. The helium atmosphere can be distinguished from a hydrogen atmosphere by its molecular compositions, i.e., the lack of \ce{CH4} and the dominance of \ce{CO}, \ce{CO2} and \ce{O2}\footnote{The transmission spectra of helium atmospheres can still have prominent \ce{H2O} features because \ce{H2O} absorption is so strong that a small mixing ratio can lead to large features.} (see Figure \ref{Transmission}). These gases have spectral features in the visible and near-infrared wavelengths. The helium atmosphere, with a mean molecular mass of $\sim4$, has a large scale height for the spectral features to manifest strongly in transmission spectra \citep{MillerRicci2009}. In fact, in comparison with recently acquired transmission spectra of Neptune- and sub-Neptune-sized exoplanets \citep[e.g.][]{Bean2010,Berta2012,Kreidberg2014,Fraine2014,Knutson2014,Knutson2014b}, observing helium atmospheres is within the reach of current observatories, amid complication of clouds and aerosols masking the molecular features \citep[e.g.][]{Benneke2013,Morley2013}. 

In addition to transmission spectroscopy, could helium-dominated atmospheres be directly detected? We suggest the following three methods, taking advantage of the unique physical properties of helium. First, helium has a very low heat capacity compared to hydrogen, because a helium atom has only three degrees of freedom and a hydrogen molecule can have six. For example, at 1000 K helium has a specific heat capacity of 5.2 J g$^{-1}$ K$^{-1}$, compared to the hydrogen's of 15.1 J mol$^{-1}$ K$^{-1}$ (http://webbook.nist.gov/chemistry/). This would lead to a helium-dominated atmosphere having the greatest adiabatic temperature-pressure gradient among common planetary atmospheres, given that the temperature gradient of a convective atmosphere is inversely proportional to the specific heat capacity. High-resolution infrared spectroscopy in the future could potentially measure the temperature gradient by deep spectral features, and thereby verify the dominance of helium. Second, due to its low heat capacity, a helium-dominated atmosphere is more likely to have an unshifted hot spot, i.e., a hottest zone at the substellar point, compared with a helium-dominated one. This will result in a thermal emission phase curve peaked at occultation. Third, helium is a very poor Rayleigh scatterer, with a Rayleigh scattering cross section 20 times smaller than that of \ce{H2} \citep{Tarafdar1969}. As a result, reflection of the stellar light by a helium-dominated atmosphere should be dominated by multiple scattering of clouds or aerosols in the atmosphere. This condition would lack the strong, broad linear polarization peak at a 90$^{\circ}$ phase angle \citep{Madhusudhan2012b}.

\section{Conclusion}

This paper proposes a new concept that some Neptune-sized and sub-Neptune-sized exoplanets may have fossil helium-dominated atmospheres. This concept is important because sub-Neptune-sized planets are ubiquitous in our interstellar neighborhood, and because helium atmospheres have unique molecular compositions and can be detected via transmission spectroscopy due to an extended scale hight.

The proposed helium atmospheres can be formed by atmospheric hydrodynamic escape driven by stellar irradiation. For hydrodynamic escape to produce a large impact on the overall atmospheric composition, the mass of the initial atmosphere must be $10^{-3}$ planetary mass or less, and the escape rate must be close to the diffusion-limited escape rate. How restrictive the first condition is uncertain and depends on the formation processes of Neptune- and sub-Neptune-sized planets. We show that the second condition is likely to be met by many irradiated Neptune- and sub-Neptune-sized exoplanets.

Applying the concept to the Neptune-sized planet GJ~436~b, we find that a helium atmosphere can be formed on a time scale of its age, if it has a small atmosphere envelope. Helium atmosphere models appear to provide better consistency with the planet's spectra than hydrogen atmosphere models, naturally explaining the lack of \ce{CH4} absorption and the dominance of \ce{CO} as the carbon-bearing species indicated by the thermal emission. 

The evolutionary scenarios presented in this paper are orders of magnitude in nature. Refinements are warranted for future work. In particular, we neglect the thermal evolution and the core contraction in the model, and we assume an isothermal atmospheric temperature profile for ease of calculation. These factors should be treated in full evolutionary simulations \citep[e.g.][]{Lopez2012}. We also neglect atmosphere-surface exchange. If the surface emits hydrogen at high fluxes, helium atmospheres cannot form. Therefore the evolution scenarios should also be coupled with geophysical calculations to fully map the potential outcomes of the evolution of short-period, low-mass exoplanets. Despite these caveats, we suggest that observations of Neptune- and sub-Neptune-sized exoplanets in the near future would test the atmospheric evolution theories presented here, and provide important clues on the origins of these alien worlds.

\acknowledgments
We appreciate comments on the manuscript made by members of the Yuk Yung research group at the California Institute of Technology. R.H. thanks Robert Johnson, Jeffrey Linsky, and Edwin Kite for helpful discussion. This work has utilized the MUSCLES M-dwarf UV radiation database. Support for this work was provided by NASA through Hubble Fellowship grant \#51332 awarded by the Space Telescope Science Institute, which is operated by the Association of Universities for Research in Astronomy, Inc., for NASA, under contract NAS 5-26555. Y.L.Y. and R.H. (in the later phase of this work) were supported in part by an NAI Virtual Planetary Laboratory grant NASA grant NNX09AB72G to the California Institute of Technology. Part of the research was carried out at the Jet Propulsion Laboratory, California Institute of Technology, under a contract with the National Aeronautics and Space Administration.

\appendix

\section{A Broad-band Emission Spectrum of GJ~436~b Obtained by {\it Spitzer}}

\label{Observation}

A broad-band emission spectrum of GJ~436~b in 3-30 $\mu$m was obtained by measuring the eclipse depths as the planet occults its parent star by {\it Spitzer} \citep{Stevenson2010}. Recently, the same observations were re-analyzed by an independent group, and new values of the eclipse depths were reported \citep{Lanotte2014}. The two datasets are consistent in the detection of an emission flux at 3.6 $\mu$m, and the non-detection at 4.5 $\mu$m. However, the specific values and uncertainties for the emission fluxes are different between the two datasets. The more recent dataset has a significantly smaller eclipse depth at 3.6 $\mu$m, and slightly smaller eclipse depths at 5.8 $\mu$m and 8.0 $\mu$m.
  
One should exercise caution when interpreting the reported values, as the multiple-wavelength observations of Neptune-sized exoplanets have only started and our understandings of instrumental systematic errors and our data analysis technique are still improving. The fidelity of the data may be best established by repeating the measurements and verifying that the results are consistent within the quoted errors. For GJ~436~b, this has only been done for the 4.5-$\mu$m and 8.0-$\mu$m bands. The 8-$\mu$m emission flux has been re-observed multiple times and the results of these repeated observations are consistent with each other at $1\sigma$ level \citep{Stevenson2010,Knutson2011}. 

Unfortunately, the measurements of the eclipse depth at 3.6 $\mu$m have not converged. The more recent analysis claimed that the 3.6-$\mu$m light curve of GJ~436 of \cite{Stevenson2010} exhibits significant variability out-of-eclipse, which may be due to incomplete removal of systematic errors. The more recent analysis appears to be able to remove this variability, but derive a much smaller eclipse depth. However, this analysis used ordinary aperture photometry assuming small apertures \citep{Lanotte2014}. It is therefore unclear how well the intrapixel sensitivity variations have been corrected in their analysis \citep{Stevenson2012}. Repeating the measurements of the eclipse depth of GJ~436~b at 3.6 $\mu$m would be necessary to pinpoint its emission flux.

\section{Model-Observation Discrepancy at 16-micron}

\label{Residual}

A remaining challenge for GJ~436~b is the high thermal emission flux of the planet detected at 16 $\mu$m, implying little \ce{CO2} absorption in the atmosphere. Our best-fit models differ from the observation at this wavelength by 2.5 $\sigma$, and this is the dominant source of $\chi^2$ for the best-fit models (Figure \ref{Model}). We have made several attempts to improve the fit.

First, making the carbon to oxygen ratio unity or very close to unity does not solve the problem. Although the amount of \ce{CO2} can be decreased, the amount of \ce{CH4} will be increased for a carbon to oxygen ratio closer to unity. As the 3.6-$\mu$m measurement for \ce{CH4} has a much smaller error bar than the 16-$\mu$m measurement for \ce{CO2}, our fitting still favors a solution that primarily satisfies the 3.6-$\mu$m constraint. Here we note that further increasing the internal heat flux may suppress \ce{CH4} for large carbon to oxygen ratios, and such a large heat flux might be physically plausible if tidal heating is efficient \citep{Agundez2014}.

Second, we find that modifying the temperature profile will improve the apparent goodness of fit. The results shown in this paper are based on grey-atmosphere temperature profiles with the opacities consistent with the composition. The calculation has also assumed a dry adiabatic lapse rate of helium for the irradiation-induced convective layer that the thermal emission spectrum is probing. The dry adiabatic lapse rate of helium is quite large, which amplifies the size of spectral features including the \ce{CO2} absorption feature. As a test we have altered the temperature-pressure profile gradient for pressures smaller than 0.4 bar, mimicking an adiabatic lapse rate softened by condensable species, or an inefficient day-night heat recirculation \citep{Lewis2010}. We found that the fit to the 16-$\mu$m emission flux can be improved to within 2 $\sigma$ without altering the composition. 

\section{Evolution History of X-ray and EUV Fluxes}

\label{Xray}

The hydrodynamic escape is driven by of X-ray and EUV radiation from the star. Although the stellar X-ray flux can be measured, the EUV flux is not directly observable due to strong interstellar absorption. For GJ~436, its EUV flux may be estimated either from the X-ray flux or from the Ly-$\alpha$ flux.

The X-ray flux of GJ~436 has been measured to be $\log(L_{\rm X-ray}) =20.16$ \citep{Poppenhaeger2010} or $\log(L_{\rm X-ray}) =18.96$ \citep{Sanz-Forcada2011}, in the unit of J s$^{-1}$. Using a fitting relationship between the X-ray flux and the EUV flux \citep{Sanz-Forcada2011}, we estimate an EUV flux of $\log(L_{\rm EUV})=20.13$, but this value could vary by 3 orders of magnitude due to uncertainties in the fitting relationship.

Alternatively, one could estimate the EUV flux from the Ly-$\alpha$ flux. Reconstruction of the Ly-$\alpha$ flux gives an estimate of $\log(L_{\rm Ly-\alpha})=20.65$ \citep{Ehrenreich2011, France2013}. Using fitting relationships between the fluxes of multiple EUV bands and the Ly-$\alpha$ flux \citep{Linsky2014}, we estimate $\log(L_{\rm EUV})=20.54$, which is consistent with the estimate from the X-ray flux.

Since we are concerned about the evolution of the atmosphere on GJ~436~b, an age-dependent evolutionary history of the X-ray and the EUV fluxes is necessary. We adopt the following empirical estimates for stellar EUV fluxes based on X-ray observations and coronal modeling of 80 stars with spectral types spanning from M to F \citep{Sanz-Forcada2011}
\begin{eqnarray}
&&\log(L_{\rm EUV}) = 22.12-1.24\log(\tau),\\
&&\log(L_{\rm X-ray}) = 21.28-1.44\log(\tau),
\end{eqnarray}
where the luminosity has a unit of J s$^{-1}$ and the age ($\tau$) has a unit of billion year. As a fact check, this adopted evolution history gives current fluxes of $\log(L_{\rm X-ray}) =19.84$ and $\log(L_{\rm EUV}) =20.88$, and these values are well within the ranges permitted by the above-mentioned observations. The age dependency is also consistent with a recent study \citep{Shkolnik2014}.

\bibliographystyle{apj}
\bibliography{master}

\newcommand{\noopsort}[1]{} \newcommand{\printfirst}[2]{#1}
  \newcommand{\singleletter}[1]{#1} \newcommand{\switchargs}[2]{#2#1}
\begin{thebibliography}{}
\expandafter\ifx\csname natexlab\endcsname\relax\def\natexlab#1{#1}\fi

\bibitem[{{Ag{\'u}ndez} {et~al.}(2014){Ag{\'u}ndez}, {Venot}, {Selsis}, \&
  {Iro}}]{Agundez2014}
{Ag{\'u}ndez}, M., {Venot}, O., {Selsis}, F., \& {Iro}, N. 2014, Astrophysical
  Journal, 781, 68

\bibitem[{{Allard} {et~al.}(1997){Allard}, {Hauschildt}, {Alexander}, \&
  {Starrfield}}]{Allard1997}
{Allard}, F., {Hauschildt}, P.~H., {Alexander}, D.~R., \& {Starrfield}, S.
  1997, Annual Review of Astronomy \& Astrophysics, 35, 137

\bibitem[{{Alonso} {et~al.}(2008){Alonso}, {Barbieri}, {Rabus}, {Deeg},
  {Belmonte}, \& {Almenara}}]{Alonso2008}
{Alonso}, R., {Barbieri}, M., {Rabus}, M., {et~al.} 2008, Astronomy \&
  Astrophysics, 487, L5

\bibitem[{{Ballard} {et~al.}(2010){Ballard}, {Christiansen}, {Charbonneau},
  {Deming}, {Holman}, {Fabrycky}, {A'Hearn}, {Wellnitz}, {Barry}, {Kuchner},
  {Livengood}, {Hewagama}, {Sunshine}, {Hampton}, {Lisse}, {Seager}, \&
  {Veverka}}]{Ballard2010}
{Ballard}, S., {Christiansen}, J.~L., {Charbonneau}, D., {et~al.} 2010,
  Astrophysical Journal, 716, 1047

\bibitem[{{Baraffe} {et~al.}(2008){Baraffe}, {Chabrier}, \&
  {Barman}}]{Baraffe2008}
{Baraffe}, I., {Chabrier}, G., \& {Barman}, T. 2008, Astronomy \& Astrophysics,
  482, 315

\bibitem[{{Barman} {et~al.}(2005){Barman}, {Hauschildt}, \&
  {Allard}}]{Barman2005}
{Barman}, T.~S., {Hauschildt}, P.~H., \& {Allard}, F. 2005, Astrophysical
  Journal, 632, 1132

\bibitem[{{Bean} {et~al.}(2010){Bean}, {Miller-Ricci Kempton}, \&
  {Homeier}}]{Bean2010}
{Bean}, J.~L., {Miller-Ricci Kempton}, E., \& {Homeier}, D. 2010, Nature, 468,
  669

\bibitem[{{Beaulieu} {et~al.}(2011){Beaulieu}, {Tinetti}, {Kipping}, {Ribas},
  {Barber}, {Cho}, {Polichtchouk}, {Tennyson}, {Yurchenko}, {Griffith},
  {Batista}, {Waldmann}, {Miller}, {Carey}, {Mousis}, {Fossey}, \&
  {Aylward}}]{Beaulieu2011}
{Beaulieu}, J.-P., {Tinetti}, G., {Kipping}, D.~M., {et~al.} 2011,
  Astrophysical Journal, 731, 16

\bibitem[{{Ben-Jaffel} \& {Sona Hosseini}(2010)}]{BenJaffel2010}
{Ben-Jaffel}, L., \& {Sona Hosseini}, S. 2010, The Astrophysical Journal, 709,
  1284

\bibitem[{{Benneke} \& {Seager}(2013)}]{Benneke2013}
{Benneke}, B., \& {Seager}, S. 2013, The Astrophysical Journal, 778, 153

\bibitem[{{Berta} {et~al.}(2012){Berta}, {Charbonneau}, {D{\'e}sert},
  {Miller-Ricci Kempton}, {McCullough}, {Burke}, {Fortney}, {Irwin}, {Nutzman},
  \& {Homeier}}]{Berta2012}
{Berta}, Z.~K., {Charbonneau}, D., {D{\'e}sert}, J.-M., {et~al.} 2012,
  Astrophysical Journal, 747, 35

\bibitem[{{Borysow} {et~al.}(1988){Borysow}, {Frommhold}, \&
  {Birnbaum}}]{Borysow1988}
{Borysow}, J., {Frommhold}, L., \& {Birnbaum}, G. 1988, Astrophysical Journal,
  326, 509

\bibitem[{{Burrows} \& {Sharp}(1999)}]{Burrows1999}
{Burrows}, A., \& {Sharp}, C.~M. 1999, Astrophysical Journal, 512, 843

\bibitem[{{Butler} {et~al.}(2004){Butler}, {Vogt}, {Marcy}, {Fischer},
  {Wright}, {Henry}, {Laughlin}, \& {Lissauer}}]{Butler2004}
{Butler}, R.~P., {Vogt}, S.~S., {Marcy}, G.~W., {et~al.} 2004, Astrophysical
  Journal, 617, 580

\bibitem[{{C{\'a}ceres} {et~al.}(2009){C{\'a}ceres}, {Ivanov}, {Minniti},
  {Naef}, {Melo}, {Mason}, {Selman}, \& {Pietrzynski}}]{Caceres2009}
{C{\'a}ceres}, C., {Ivanov}, V.~D., {Minniti}, D., {et~al.} 2009, Astronomy \&
  Astrophysics, 507, 481

\bibitem[{{Cecchi-Pestellini} {et~al.}(2006){Cecchi-Pestellini}, {Ciaravella},
  \& {Micela}}]{CecchiPestellini2006}
{Cecchi-Pestellini}, C., {Ciaravella}, A., \& {Micela}, G. 2006, Astronomy \&
  Astrophysics, 458, L13

\bibitem[{{Cecchi-Pestellini} {et~al.}(2009){Cecchi-Pestellini}, {Ciaravella},
  {Micela}, \& {Penz}}]{CecchiPestellini2009}
{Cecchi-Pestellini}, C., {Ciaravella}, A., {Micela}, G., \& {Penz}, T. 2009,
  Astronomy \& Astrophysics, 496, 863

\bibitem[{{Deming} {et~al.}(2007){Deming}, {Harrington}, {Laughlin}, {Seager},
  {Navarro}, {Bowman}, \& {Horning}}]{Deming2007}
{Deming}, D., {Harrington}, J., {Laughlin}, G., {et~al.} 2007, Astrophysical
  Journal Letters, 667, L199

\bibitem[{{Demory} {et~al.}(2007){Demory}, {Gillon}, {Barman}, {Bonfils},
  {Mayor}, {Mazeh}, {Queloz}, {Udry}, {Bouchy}, {Delfosse}, {Forveille},
  {Mallmann}, {Pepe}, \& {Perrier}}]{Demory2007}
{Demory}, B.-O., {Gillon}, M., {Barman}, T., {et~al.} 2007, Astronomy \&
  Astrophysics, 475, 1125

\bibitem[{{Ehrenreich} {et~al.}(2011){Ehrenreich}, {Lecavelier Des Etangs}, \&
  {Delfosse}}]{Ehrenreich2011}
{Ehrenreich}, D., {Lecavelier Des Etangs}, A., \& {Delfosse}, X. 2011,
  Astronomy \& Astrophysics, 529, A80

\bibitem[{{Erkaev} {et~al.}(2007){Erkaev}, {Kulikov}, {Lammer}, {Selsis},
  {Langmayr}, {Jaritz}, \& {Biernat}}]{Erkaev2007}
{Erkaev}, N.~V., {Kulikov}, Y.~N., {Lammer}, H., {et~al.} 2007, Astronomy \&
  Astrophysics, 472, 329

\bibitem[{{Erwin} {et~al.}(2013){Erwin}, {Tucker}, \& {Johnson}}]{Erwin2013}
{Erwin}, J., {Tucker}, O.~J., \& {Johnson}, R.~E. 2013, Icarus, 226, 375

\bibitem[{{Fortney} {et~al.}(2013){Fortney}, {Mordasini}, {Nettelmann},
  {Kempton}, {Greene}, \& {Zahnle}}]{Fortney2013}
{Fortney}, J.~J., {Mordasini}, C., {Nettelmann}, N., {et~al.} 2013,
  Astrophysical Journal, 775, 80

\bibitem[{{Fraine} {et~al.}(2014){Fraine}, {Deming}, {Benneke}, {Knutson},
  {Jord{\'a}n}, {Espinoza}, {Madhusudhan}, {Wilkins}, \&
  {Todorov}}]{Fraine2014}
{Fraine}, J., {Deming}, D., {Benneke}, B., {et~al.} 2014, Nature, 513, 526

\bibitem[{{France} {et~al.}(2013){France}, {Froning}, {Linsky}, {Roberge},
  {Stocke}, {Tian}, {Bushinsky}, {D{\'e}sert}, {Mauas}, {Vieytes}, \&
  {Walkowicz}}]{France2013}
{France}, K., {Froning}, C.~S., {Linsky}, J.~L., {et~al.} 2013, Astrophysical
  Journal, 763, 149

\bibitem[{{Fressin} {et~al.}(2013){Fressin}, {Torres}, {Charbonneau}, {Bryson},
  {Christiansen}, {Dressing}, {Jenkins}, {Walkowicz}, \&
  {Batalha}}]{Fressin2013}
{Fressin}, F., {Torres}, G., {Charbonneau}, D., {et~al.} 2013, Astrophysical
  Journal, 766, 81

\bibitem[{{Garc{\'{\i}}a Mu{\~n}oz}(2007)}]{GarciaMunoz2007}
{Garc{\'{\i}}a Mu{\~n}oz}, A. 2007, Planetary and Space Science, 55, 1426

\bibitem[{{Gierasch} \& {Conrath}(1985)}]{Gierasch1985}
{Gierasch}, P.~J., \& {Conrath}, B.~J. 1985, {Energy conversion processes in
  the outer planets} (Cambridge University Press, Cambridge and New York),
  121--146

\bibitem[{{Gillon} {et~al.}(2007){Gillon}, {Pont}, {Demory}, {Mallmann},
  {Mayor}, {Mazeh}, {Queloz}, {Shporer}, {Udry}, \& {Vuissoz}}]{Gillon2007}
{Gillon}, M., {Pont}, F., {Demory}, B.-O., {et~al.} 2007, Astronomy \&
  Astrophysics, 472, L13

\bibitem[{{Guo}(2013)}]{Guo2013}
{Guo}, J.~H. 2013, The Astrophysical Journal, 766, 102

\bibitem[{{Hansen} \& {Murray}(2012)}]{Hansen2012}
{Hansen}, B.~M.~S., \& {Murray}, N. 2012, Astrophysical Journal, 751, 158

\bibitem[{{Holmstr{\"o}m} {et~al.}(2008){Holmstr{\"o}m}, {Ekenb{\"a}ck},
  {Selsis}, {Penz}, {Lammer}, \& {Wurz}}]{Holmstrom2008}
{Holmstr{\"o}m}, M., {Ekenb{\"a}ck}, A., {Selsis}, F., {et~al.} 2008, Nature,
  451, 970

\bibitem[{{Howard}(2013)}]{Howard2013}
{Howard}, A.~W. 2013, Science, 340, 572

\bibitem[{{Hu} \& {Seager}(2014)}]{Hu2014}
{Hu}, R., \& {Seager}, S. 2014, Astrophysical Journal, 784, 63

\bibitem[{{Hu} {et~al.}(2012){Hu}, {Seager}, \& {Bains}}]{Hu2012}
{Hu}, R., {Seager}, S., \& {Bains}, W. 2012, Astrophysical Journal, 761, 166

\bibitem[{{Hu} {et~al.}(2013){Hu}, {Seager}, \& {Bains}}]{Hu2013}
---. 2013, Astrophysical Journal, 769, 6

\bibitem[{{Hunten} {et~al.}(1987){Hunten}, {Pepin}, \& {Walker}}]{Hunten1987}
{Hunten}, D.~M., {Pepin}, R.~O., \& {Walker}, J.~C.~G. 1987, Icarus, 69, 532

\bibitem[{{Ikoma} \& {Hori}(2012)}]{Ikoma2012}
{Ikoma}, M., \& {Hori}, Y. 2012, Astrophysical Journal, 753, 66

\bibitem[{{Inamdar} \& {Schlichting}(2014)}]{Inamdar2014}
{Inamdar}, N.~K., \& {Schlichting}, H.~E. 2014, ArXiv e-prints, arXiv:1412.4440

\bibitem[{{Johnson} {et~al.}(2013){Johnson}, {Volkov}, \&
  {Erwin}}]{Johnson2013}
{Johnson}, R.~E., {Volkov}, A.~N., \& {Erwin}, J.~T. 2013, Astrophysical
  Journal Letters, 768, L4

\bibitem[{Kislyakova {et~al.}(2014)Kislyakova, Holmstr{\"o}m, Lammer, Odert, \&
  Khodachenko}]{Kislyakova2014}
Kislyakova, K.~G., Holmstr{\"o}m, M., Lammer, H., Odert, P., \& Khodachenko,
  M.~L. 2014, Science, 346, 981

\bibitem[{{Knutson} {et~al.}(2014{\natexlab{a}}){Knutson}, {Benneke}, {Deming},
  \& {Homeier}}]{Knutson2014}
{Knutson}, H.~A., {Benneke}, B., {Deming}, D., \& {Homeier}, D.
  2014{\natexlab{a}}, Nature, 505, 66

\bibitem[{{Knutson} {et~al.}(2011){Knutson}, {Madhusudhan}, {Cowan},
  {Christiansen}, {Agol}, {Deming}, {D{\'e}sert}, {Charbonneau}, {Henry},
  {Homeier}, {Langton}, {Laughlin}, \& {Seager}}]{Knutson2011}
{Knutson}, H.~A., {Madhusudhan}, N., {Cowan}, N.~B., {et~al.} 2011,
  Astrophysical Journal, 735, 27

\bibitem[{{Knutson} {et~al.}(2014{\natexlab{b}}){Knutson}, {Dragomir},
  {Kreidberg}, {Kempton}, {McCullough}, {Fortney}, {Bean}, {Gillon}, {Homeier},
  \& {Howard}}]{Knutson2014b}
{Knutson}, H.~A., {Dragomir}, D., {Kreidberg}, L., {et~al.} 2014{\natexlab{b}},
  The Astrophysical Journal, 794, 155

\bibitem[{Koskinen {et~al.}(2013)Koskinen, Harris, Yelle, \&
  Lavvas}]{Koskinen2013}
Koskinen, T., Harris, M., Yelle, R., \& Lavvas, P. 2013, Icarus, 226, 1678

\bibitem[{{Koskinen} {et~al.}(2010){Koskinen}, {Yelle}, {Lavvas}, \&
  {Lewis}}]{Koskinen2010}
{Koskinen}, T.~T., {Yelle}, R.~V., {Lavvas}, P., \& {Lewis}, N.~K. 2010, The
  Astrophysical Journal, 723, 116

\bibitem[{{Kreidberg} {et~al.}(2014){Kreidberg}, {Bean}, {D{\'e}sert},
  {Benneke}, {Deming}, {Stevenson}, {Seager}, {Berta-Thompson}, {Seifahrt}, \&
  {Homeier}}]{Kreidberg2014}
{Kreidberg}, L., {Bean}, J.~L., {D{\'e}sert}, J.-M., {et~al.} 2014, Nature,
  505, 69

\bibitem[{{Kulow} {et~al.}(2014){Kulow}, {France}, {Linsky}, \&
  {Loyd}}]{Kulow2014}
{Kulow}, J.~R., {France}, K., {Linsky}, J., \& {Loyd}, R.~O.~P. 2014,
  Astrophysical Journal, 786, 132

\bibitem[{{Lammer} {et~al.}(2013){Lammer}, {Erkaev}, {Odert}, {Kislyakova},
  {Leitzinger}, \& {Khodachenko}}]{Lammer2013}
{Lammer}, H., {Erkaev}, N.~V., {Odert}, P., {et~al.} 2013, Monthly Notices of
  the Royal Astronomical Society, 430, 1247

\bibitem[{{Lammer} {et~al.}(2003){Lammer}, {Selsis}, {Ribas}, {Guinan},
  {Bauer}, \& {Weiss}}]{Lammer2003}
{Lammer}, H., {Selsis}, F., {Ribas}, I., {et~al.} 2003, Astrophysical Journal
  Letters, 598, L121

\bibitem[{{Lanotte} {et~al.}(2014){Lanotte}, {Gillon}, {Demory}, {Fortney},
  {Astudillo}, {Bonfils}, \& {Magain}}]{Lanotte2014}
{Lanotte}, A.~A., {Gillon}, M., {Demory}, B.-O., {et~al.} 2014, ArXiv e-prints,
  arXiv:1409.4038

\bibitem[{{Lecavelier des Etangs} {et~al.}(2004){Lecavelier des Etangs},
  {Vidal-Madjar}, {McConnell}, \& {H{\'e}brard}}]{Lecavelier2004}
{Lecavelier des Etangs}, A., {Vidal-Madjar}, A., {McConnell}, J.~C., \&
  {H{\'e}brard}, G. 2004, Astronomy and Astrophysics, 418, L1

\bibitem[{{Lewis} {et~al.}(2010){Lewis}, {Showman}, {Fortney}, {Marley},
  {Freedman}, \& {Lodders}}]{Lewis2010}
{Lewis}, N.~K., {Showman}, A.~P., {Fortney}, J.~J., {et~al.} 2010,
  Astrophysical Journal, 720, 344

\bibitem[{{Liang} {et~al.}(2003){Liang}, {Parkinson}, {Lee}, {Yung}, \&
  {Seager}}]{Liang2003}
{Liang}, M.-C., {Parkinson}, C.~D., {Lee}, A.~Y.-T., {Yung}, Y.~L., \&
  {Seager}, S. 2003, Astrophysical Journal Letters, 596, L247

\bibitem[{{Line} {et~al.}(2014){Line}, {Knutson}, {Wolf}, \& {Yung}}]{Line2014}
{Line}, M.~R., {Knutson}, H., {Wolf}, A.~S., \& {Yung}, Y.~L. 2014,
  Astrophysical Journal, 783, 70

\bibitem[{{Line} {et~al.}(2011){Line}, {Vasisht}, {Chen}, {Angerhausen}, \&
  {Yung}}]{Line2011}
{Line}, M.~R., {Vasisht}, G., {Chen}, P., {Angerhausen}, D., \& {Yung}, Y.~L.
  2011, Astrophysical Journal, 738, 32

\bibitem[{{Linsky} {et~al.}(2014){Linsky}, {Fontenla}, \&
  {France}}]{Linsky2014}
{Linsky}, J.~L., {Fontenla}, J., \& {France}, K. 2014, Astrophysical Journal,
  780, 61

\bibitem[{{Llama} \& {Shkolnik}(2015)}]{Llama2015}
{Llama}, J., \& {Shkolnik}, E.~L. 2015, The Astrophysical Journal, 802, 41

\bibitem[{{Lopez} {et~al.}(2012){Lopez}, {Fortney}, \& {Miller}}]{Lopez2012}
{Lopez}, E.~D., {Fortney}, J.~J., \& {Miller}, N. 2012, Astrophysical Journal,
  761, 59

\bibitem[{{Loyd} \& {France}(2014)}]{Loyd2014}
{Loyd}, R.~O.~P., \& {France}, K. 2014, The Astrophysical Journal Supplement,
  211, 9

\bibitem[{{Madhusudhan} \& {Burrows}(2012)}]{Madhusudhan2012b}
{Madhusudhan}, N., \& {Burrows}, A. 2012, Astrophysical Journal, 747, 25

\bibitem[{{Madhusudhan} \& {Seager}(2011)}]{Madhusudhan2011}
{Madhusudhan}, N., \& {Seager}, S. 2011, Astrophysical Journal, 729, 41

\bibitem[{{Mason} \& {Marrero}(1970)}]{Mason1970}
{Mason}, E.~A., \& {Marrero}, T.~R. 1970, Advances in Atomic and Molecular
  Physics, 6, 155

\bibitem[{{Miller-Ricci} {et~al.}(2009){Miller-Ricci}, {Seager}, \&
  {Sasselov}}]{MillerRicci2009}
{Miller-Ricci}, E., {Seager}, S., \& {Sasselov}, D. 2009, Astrophysical
  Journal, 690, 1056

\bibitem[{{Morello} {et~al.}(2015){Morello}, {Waldmann}, {Tinetti}, {Howarth},
  {Micela}, \& {Allard}}]{Morello2015}
{Morello}, G., {Waldmann}, I.~P., {Tinetti}, G., {et~al.} 2015, ArXiv e-prints,
  arXiv:1501.05866

\bibitem[{{Morley} {et~al.}(2013){Morley}, {Fortney}, {Kempton}, {Marley},
  {Visscher}, \& {Zahnle}}]{Morley2013}
{Morley}, C.~V., {Fortney}, J.~J., {Kempton}, E.~M.-R., {et~al.} 2013,
  Astrophysical Journal, 775, 33

\bibitem[{{Moses} {et~al.}(2011){Moses}, {Visscher}, {Fortney}, {Showman},
  {Lewis}, {Griffith}, {Klippenstein}, {Shabram}, {Friedson}, {Marley}, \&
  {Freedman}}]{Moses2011}
{Moses}, J.~I., {Visscher}, C., {Fortney}, J.~J., {et~al.} 2011, Astrophysical
  Journal, 737, 15

\bibitem[{{Moses} {et~al.}(2013){Moses}, {Line}, {Visscher}, {Richardson},
  {Nettelmann}, {Fortney}, {Barman}, {Stevenson}, \&
  {Madhusudhan}}]{Moses2013a}
{Moses}, J.~I., {Line}, M.~R., {Visscher}, C., {et~al.} 2013, Astrophysical
  Journal, 777, 34

\bibitem[{{Murray-Clay} {et~al.}(2009){Murray-Clay}, {Chiang}, \&
  {Murray}}]{Murray-Clay2009}
{Murray-Clay}, R.~A., {Chiang}, E.~I., \& {Murray}, N. 2009, Astrophysical
  Journal, 693, 23

\bibitem[{{Nettelmann} {et~al.}(2010){Nettelmann}, {Kramm}, {Redmer}, \&
  {Neuh{\"a}user}}]{Nettelmann2010}
{Nettelmann}, N., {Kramm}, U., {Redmer}, R., \& {Neuh{\"a}user}, R. 2010,
  Astronomy \& Astrophysics, 523, A26

\bibitem[{{Owen} \& {Jackson}(2012)}]{Owen2012}
{Owen}, J.~E., \& {Jackson}, A.~P. 2012, Monthly Notices of the Royal
  Astronomical Society, 425, 2931

\bibitem[{{Owen} \& {Wu}(2013)}]{Owen2013}
{Owen}, J.~E., \& {Wu}, Y. 2013, Astrophysical Journal, 775, 105

\bibitem[{{Parmentier} {et~al.}(2013){Parmentier}, {Showman}, \&
  {Lian}}]{Parmentier2013}
{Parmentier}, V., {Showman}, A.~P., \& {Lian}, Y. 2013, Astronomy \&
  Astrophysics, 558, A91

\bibitem[{{Pont} {et~al.}(2009){Pont}, {Gilliland}, {Knutson}, {Holman}, \&
  {Charbonneau}}]{Pont2009}
{Pont}, F., {Gilliland}, R.~L., {Knutson}, H., {Holman}, M., \& {Charbonneau},
  D. 2009, Monthly Notices of the Royal Astronomical Society, 393, L6

\bibitem[{{Poppenhaeger} {et~al.}(2010){Poppenhaeger}, {Robrade}, \&
  {Schmitt}}]{Poppenhaeger2010}
{Poppenhaeger}, K., {Robrade}, J., \& {Schmitt}, J.~H.~M.~M. 2010, Astronomy \&
  Astrophysics, 515, A98

\bibitem[{{Rafikov}(2006)}]{Rafikov2006}
{Rafikov}, R.~R. 2006, Astrophysical Journal, 648, 666

\bibitem[{{Rogers} \& {Seager}(2010)}]{Rogers2010}
{Rogers}, L.~A., \& {Seager}, S. 2010, Astrophysical Journal, 712, 974

\bibitem[{{Sanz-Forcada} {et~al.}(2011){Sanz-Forcada}, {Micela}, {Ribas},
  {Pollock}, {Eiroa}, {Velasco}, {Solano}, \&
  {Garc{\'{\i}}a-{\'A}lvarez}}]{Sanz-Forcada2011}
{Sanz-Forcada}, J., {Micela}, G., {Ribas}, I., {et~al.} 2011, Astronomy \&
  Astrophysics, 532, A6

\bibitem[{Schunk \& Nagy(1980)}]{Schunk1980}
Schunk, R.~W., \& Nagy, A.~F. 1980, Reviews of Geophysics, 18, 813

\bibitem[{{Seager} \& {Deming}(2010)}]{seager2010b}
{Seager}, S., \& {Deming}, D. 2010, Annual Review of Astronomy \& Astrophysics,
  48, 631

\bibitem[{{Shkolnik} \& {Barman}(2014)}]{Shkolnik2014}
{Shkolnik}, E.~L., \& {Barman}, T.~S. 2014, ArXiv e-prints, arXiv:1407.1344

\bibitem[{{Spitzer}(1978)}]{Spitzer1978}
{Spitzer}, L. 1978, {Physical processes in the interstellar medium} (New York
  Wiley-Interscience)

\bibitem[{{Stevenson} {et~al.}(2010){Stevenson}, {Harrington}, {Nymeyer},
  {Madhusudhan}, {Seager}, {Bowman}, {Hardy}, {Deming}, {Rauscher}, \&
  {Lust}}]{Stevenson2010}
{Stevenson}, K.~B., {Harrington}, J., {Nymeyer}, S., {et~al.} 2010, Nature,
  464, 1161

\bibitem[{{Stevenson} {et~al.}(2012){Stevenson}, {Harrington}, {Lust}, {Lewis},
  {Montagnier}, {Moses}, {Visscher}, {Blecic}, {Hardy}, {Cubillos}, \&
  {Campo}}]{Stevenson2012}
{Stevenson}, K.~B., {Harrington}, J., {Lust}, N.~B., {et~al.} 2012,
  Astrophysical Journal, 755, 9

\bibitem[{{Sudarsky} {et~al.}(2003){Sudarsky}, {Burrows}, \&
  {Hubeny}}]{Sudarsky2003}
{Sudarsky}, D., {Burrows}, A., \& {Hubeny}, I. 2003, Astrophysical Journal,
  588, 1121

\bibitem[{{Tarafdar} \& {Vardya}(1969)}]{Tarafdar1969}
{Tarafdar}, S.~P., \& {Vardya}, M.~S. 1969, Monthly Notices of the Royal
  Astronomical Society, 145, 171

\bibitem[{{Tian} {et~al.}(2005){Tian}, {Toon}, {Pavlov}, \& {De
  Sterck}}]{Tian2005}
{Tian}, F., {Toon}, O.~B., {Pavlov}, A.~A., \& {De Sterck}, H. 2005,
  Astrophysical Journal, 621, 1049

\bibitem[{{Torres} {et~al.}(2008){Torres}, {Winn}, \& {Holman}}]{Torres2008}
{Torres}, G., {Winn}, J.~N., \& {Holman}, M.~J. 2008, Astrophysical Journal,
  677, 1324

\bibitem[{{Tucker} {et~al.}(2012){Tucker}, {Erwin}, {Deighan}, {Volkov}, \&
  {Johnson}}]{Tucker2012}
{Tucker}, O.~J., {Erwin}, J.~T., {Deighan}, J.~I., {Volkov}, A.~N., \&
  {Johnson}, R.~E. 2012, Icarus, 217, 408

\bibitem[{{Volkov} \& {Johnson}(2013)}]{Volkov2013}
{Volkov}, A.~N., \& {Johnson}, R.~E. 2013, Astrophysical Journal, 765, 90

\bibitem[{{Yelle}(2004)}]{Yelle2004}
{Yelle}, R.~V. 2004, Icarus, 170, 167

\bibitem[{{Yung} \& {Demore}(1999)}]{Yung1999}
{Yung}, Y.~L., \& {Demore}, W.~B., eds. 1999, {Photochemistry of planetary
  atmospheres}

\bibitem[{{Zahnle} {et~al.}(1990){Zahnle}, {Kasting}, \&
  {Pollack}}]{Zahnle1990}
{Zahnle}, K., {Kasting}, J.~F., \& {Pollack}, J.~B. 1990, Icarus, 84, 502

\bibitem[{{Zahnle} \& {Kasting}(1986)}]{Zahnle1986a}
{Zahnle}, K.~J., \& {Kasting}, J.~F. 1986, Icarus, 68, 462

\end{thebibliography}

\end{document}